\documentclass{emulateapj}
\usepackage{times}
\usepackage{graphicx}

\shorttitle{Black Widow Light Curves}
\shortauthors{Draghis et al.}

\begin{document}

\title{Multiband Optical Light Curves of Black-Widow Pulsars}

\author{Paul Draghis\altaffilmark{1}, Roger W. Romani\altaffilmark{1}, Alexei V. Filippenko\altaffilmark{2,3},
Thomas G. Brink\altaffilmark{2},
WeiKang Zheng\altaffilmark{2},
Jules P. Halpern\altaffilmark{4}, \& Fernando Camilo\altaffilmark{5}}
\altaffiltext{1}{Department of Physics, Stanford University, Stanford, CA 94305-4060,
 USA; rwr@astro.stanford.edu}
\altaffiltext{2}{Department of Astronomy, University of California, Berkeley, CA 94720-3411, USA}
\altaffiltext{5}{Miller Senior Fellow, Miller Institute for Basic Research in Science, University of California, Berkeley, CA  94720, USA}
\altaffiltext{4}{Columbia Astrophysics Laboratory, Columbia University, New York, NY 10027, USA}
\altaffiltext{5}{South African Radio Astronomy Observatory, Observatory 7925, South Africa}

\begin{abstract}
We collect new and archival optical observations of nine ``black-widow'' millisecond pulsar binaries.  New measurements include direct imaging with the Keck, Gemini-S, MDM, and LCO 2~m telescopes. This is supplemented by synthesized colors from Keck long-slit spectra. Four black-widow optical companions are presented here for the first time. Together these data provide multicolor photometry covering a large fraction of the orbital phase. We fit these light curves with a direct (photon) heating model using a version of the ICARUS light-curve modeling code. The fits provide distance and fill-factor estimates, inclinations, and heating powers. We compare the heating powers with the observed GeV luminosities, noting that the ratio is sensitive to pulsar distance and to the gamma-ray beaming. We make a specific correction for ``outer-gap'' model beams, but even then some sources are substantially discrepant, suggesting imperfect beaming corrections and/or errors in the fit distance.  The fits prefer large metal abundance for half of the targets, a reasonable result for these wind-stripped secondaries. The companion radii indicate substantial Roche-lobe filling, $f_c \approx 0.7-1$ except for PSR J0952$-$0607, which with $f_c< 0.5$ has a companion density $\rho \approx 10\,{\rm g\,cm^{-3}}$, suggesting unusual evolution. We note that the direct-heating fits imply large heating powers and rather small inclinations, and we speculate that unmodeled effects can introduce such bias.
\end{abstract}


\section{Introduction}

``Black-widow'' (BW) class binary pulsars lie at the extreme of the pulsar population.  They have been recycled by mass transfer in a low-mass X-ray binary (LMXB) phase and are at present evaporating the companion remnant via a relativistic pulsar wind.  For classical BWs, like the original PSR J1959+2048 system (Fruchter et al. 1988), the companion is a $\sim 0.02$--0.05\,$M_\odot$ substellar remnant of an LMXB mass donor in a $P_b\approx 2.5$--10\,hr orbit.  In the closely related ``Tidarren'' subclass \citep{rgfz16}, the remnant is $\leq 0.02\,M_\odot$ and $P_b < 2$\,hr, and the companion is the H-stripped core of a partly evolved star, evidently the descendents of the ultra-compact X-ray binaries (UCXBs). These systems are of great interest for the physics of millisecond pulsar (MSP) recycling and companion evaporation. Also, slow mass transfer during the X-ray binary phase allows substantial neutron star (NS) mass growth; some systems may even reach black hole collapse, so the heaviest BW avoiding this fate can probe the $M_{\rm max}$ NS limit. Since pulsar heating makes the sub-stellar companion visible, optical observations can probe these phenomena. For example, optical spectroscopy can be used to measure the companion composition and radial velocity. Multicolor photometry, together with light-curve modeling, plays a key role in these studies since light curve fitting constrains the distance, the companion Roche-lobe fill factor, and the heating. Most importantly, this fitting constrains the system inclination $i$, which together with spectroscopic measurements determines component masses $\propto {\rm sin}^3 i$.

We have made measurements of a set of such MSPs using a variety of telescopes, including new observations, re-analysis of archival datasets, and use of published photometry. After summarizing the targets and the new data collected (\S2), we describe our photometry  (\S3) which puts the measurements on a consistent (SDSS-based) photometric system based on differential photometry against catalog stars in the field. We have focused on $g$-, $r$-, and $i$-band imaging, with some $u$- and $z$-band measurements. Additional images, especially optical, are on the Johnson system. Direct image photometry is supplemented in a number of cases with broadband magnitudes computed from differential spectroscopy. We generally have phase overlap between the direct and synthesized photometry, helping establish the zero point (grey shift) of the synthesized points. After collecting the most complete multiband light curves possible, we fit (\S4) the data with the ICARUS light-curve synthesis code \citep{bet13}, using the common assumption that the companions are directly heated by the pulsar radiation. In \S4.2--4.5 we discuss the results and their implications for companion heating. We conclude by considering some model amendments that might produce better agreement with theoretical expectations.

\begin{figure}[t!!]
\vskip -5mm
\includegraphics[width=9cm]{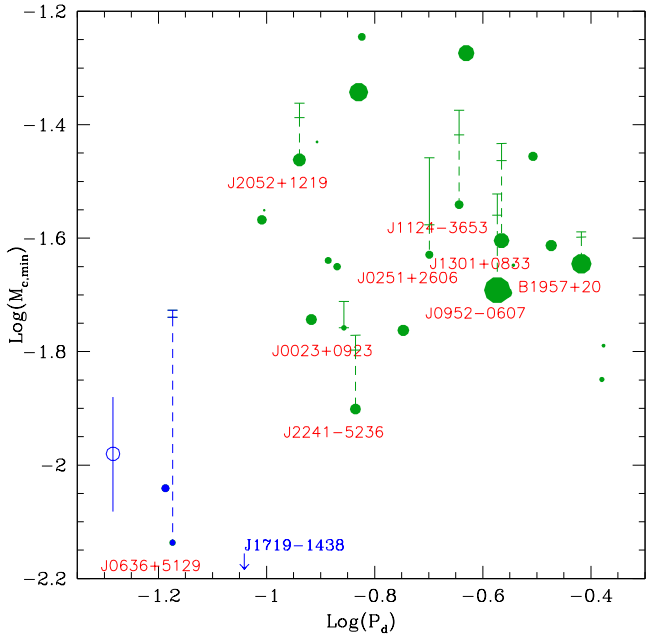}
\vskip -3mm
\caption{\label{BWpop} The black-widow region of the orbital period (in days) vs. minimum companion mass (in $M_\odot$) plane. Objects are from Lorimer's Galactic MSP compilation (http://astro.phys.wvu.edu/GalacticMSPs/GalacticMSPs.txt). The dot size is $\propto P_s^{-2}$. Short $P_b$, low $M_c$ objects at lower left (in blue, including J1653$-$0158) represent the Tidarren subclass. The objects in this paper are labeled in red. Vertical lines mark the range from $M_{c,{\rm min}}$ to the mass from the $i$ estimate of this paper (assuming $M_p = 1.5\,M_\odot$).} 
\end{figure}

\begin{figure*}[t!!]
\vskip -2mm
\hskip -1mm
\includegraphics[width=18cm]{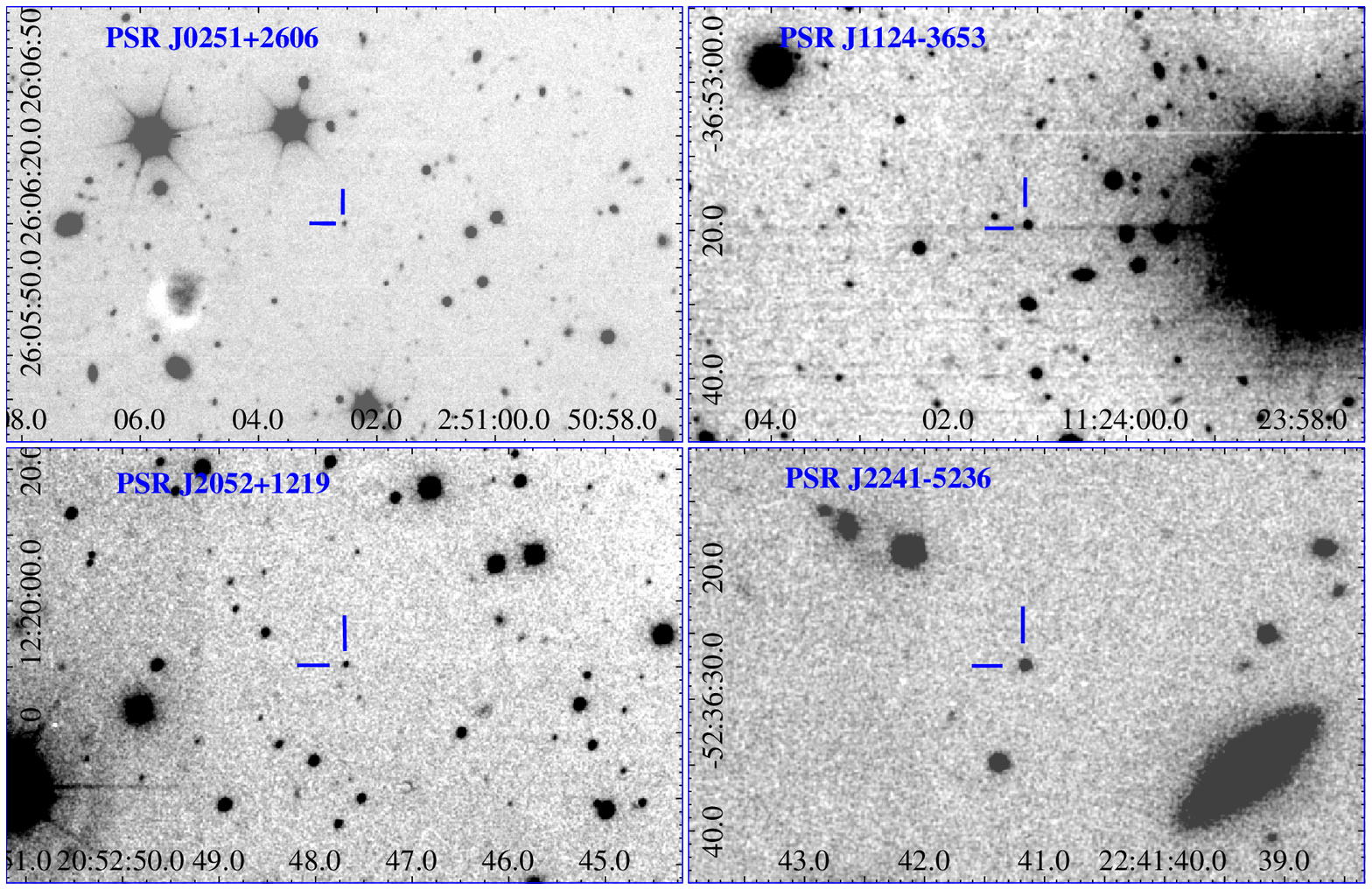}
\vskip -3mm
\caption{\label{Finder} Finder charts for four BWs near maximum brightness. J0251 -- 300\,s Keck~I/LRIS $I$; J1124 -- 300\,s GMOS-S $r$; J2052 -- 180\,s SOAR/SAMI $i$; J2241 -- 180s SOAR/SAMI $i$. All fields have N up, E to the left. } 
\end{figure*}

\section{The Black-Widow Sample}

Searches in the directions of {\it Fermi} LAT $\gamma$-ray sources have discovered over forty ``spider'' (companion-evaporating) MSPs (BWs and their ``redback'' (RB) cousins with low-mass main-sequence secondaries). This is $>10$ 
16
 times the pre-{\it Fermi} number, offering many opportunities for studying this dramatic phase in the life of interacting binaries. In Table 1 we summarize the basic observed properties of our nine systems. Figure 1 shows the BW/Tid population with our sample labeled in red.

Initial optical photometry of J0023+0923 (hereafter J0023) was published by \citet{bet13}. PSR J0636+5128's (J0636) optical counterpart was described by \citet{dr18} and \citet{kapet18}. J0952$-$0607 (J0952) had some $r$ imaging reported by \citet{baset17}. PSR J1301+0833's (J1301) counterpart was first described by \citet{lht14} with additional photometry given by \citet{rgfz16}. PSRs J0251+2606 (J0251) and J2052+1219 (J2052) were announced by \citet{cromet16}. J1124$-$3653 (J1124) was found in the GBT 350\,GHz drift-scan survey \citep{boyet13}, while J2241$-$5236 (J2241) was discovered by \citet{ket11}. To the best of our knowledge the optical counterparts of these last four BWs have not been previously described, so in Figure 2 we present finder charts showing the companions near maximum brightness.

\begin{deluxetable*}{lrrlrlrrlrr}[b!!]
\tablecaption{\label{Fits} Black Widow Parameters}
\tablehead{
\colhead{Name}&\colhead{$P_s$}&\colhead{$P_b$}&\colhead{${\dot E}_{34}$}&\colhead{$x_1$}&\colhead{$d$}&\colhead{$A_V$}&\colhead{$\mu$}&\colhead{ range} &\colhead{$f_\gamma$} &\colhead{$\Delta_\gamma$}\cr
 & ms & hr &$10^{34}$\,erg\,s$^{-1}$ &lt-s &kpc & mag & mas\,yr$^{-1}$ & $i\,{\rm mag}^a$ & $^b$ & $^b$
}
\startdata
J0023+0923   &3.05 & 3.33 & 1.60 & 0.035 & 1.3   & 0.382 & 13.9 &22.3-24.5 & 7.6 & 0\cr
J0251+2606$^e$   &2.54 & 4.86 & 1.81 & 0.066 & 1.2   & 0.382 & $>14$ &22.8-  & 4.3 & 0.50 \cr
J0636+5128   &2.87 & 1.60 & 0.58 & 0.009 & 1.1$^{c}$& 0.218 &  4.2& 22.4-23.9 & $<1$ & 0 \cr
J0952$-$0607 &1.41 & 6.42 &6.65$^d$& 0.063 & 1.7   & 0.137 & -- & 22.0-24.4 & 2.2 &  0.20 \cr
J1124$-$3653 &2.41 & 5.44 & 0.52 & 0.080 & 1.1   & 0.124 & -- &22.8-26.5  &12.5 & 0.35\cr
J1301+0833   &1.84 & 6.48 & 6.65 & 0.078 & 1.2   & 0.027 & 26.9 & 21.4-24.8 & 7.6 &0.40 \cr
J1959+2048   &1.61 & 9.17 & 16.0 & 0.089 & 1.7   & 0.600 & 30.6 &19.6-24.6 &15.3 & 0.45 \cr
J2052+1219$^e$   &1.99 & 2.75 & 3.37 & 0.061 & 3.9   & 0.328 & 15.4 & 22.0-25.1 & 5.5 & 0.35 \cr
J2241$-$5236 &2.19 & 3.50 & 2.50 & 0.026 & 1.0   & 0.040 & 17.2 &21.3-24.0&26.6 & 0.55\cr
\enddata
\tablenotetext{a}{Approximate $i$ magnitude range, when measured. For J1959, range in $R$.}
\tablenotetext{b}{4FGL {\it Fermi} LAT 0.1--100\,GeV flux in units of $10^{-12}\,{\rm erg\,cm^{-2}\,s^{-1}}$; double peak separation $\Delta_\gamma$ (0 for single peaks).}
\tablenotetext{c}{J0636 timing parallax \citep{arzet18}; all other distances are derived from the dispersion measure \citep[DM;][]{ymw17}.}
\tablenotetext{d}{From \citet{nidet19}.}
\tablenotetext{e}{J0251 and J2052 values from Deneva et al. (in prep.). Other values from ATNF catalog and astro.phys.wvu.edu/GalacticMSPs.txt. }
\end{deluxetable*}

\section {New Observations, Photometry, and Light Curves} 

\begin{deluxetable}{llll}[t!!]
\tablecaption{\label{Fits} Journal of Observations}
\tablehead{
\colhead{Name} & \colhead{Tel./Inst.}& \colhead{Exp. (s)} &  \colhead{MJD}
}
\startdata
J0023+0923 
           & LCO2m  & 4x600 $r$ & 58366.057-.078 \cr
           & & 11x600 $r$ & 58369.917-71.052 \cr
           & & 1x600 $g$ & 58372.015 \cr
           & & 3x600 $g$ & 58372.969-3.000 \cr
           & & 4x600 $g$ & 58377.968-.997 \cr
           & DEIMOS & 2x600 $R$ &  57639.456-.463 \cr
           &GMOS$^a$ & 1x620 $g$, 2x320 $i$ & 55445.506-.518 \cr
           & & 3x620 $g$, 4x320 $i$ & 55449.568-.617\cr
           &SOAR & 11x600 $i$ & 58335.875-.946 \cr
           & & 17x300 $i$ & 58336.869-.941 \cr
           & & 2x300 $g/r$ & 58336.861-.876 \cr
           \cr
           
J0251+2606 &LRIS  &200, 300 $B/R$,  & 57756.307-.939\cr
           & & 200, 6x300, 2x450 $V$, & \cr
           & & 200, 3x300, 4x450 $I$ & \cr
           & &13x320 $g$, &57991.566-.621 \cr
           & & 6x300 $R$, 7x300 $I$ & \cr
           \cr
J0636+5128 & GMOS$^a$&10x420 $g$, 9x420 $r$ & 57012.324-.420 \cr
           & NIRI$^a$ &21x60 $K$         &56966.588-.610\cr
           &          &25x60 $K$         &56999.422-.447\cr
           & NIRC2$^a$ & 27x60 $H$, 28x60 $K'$ & 56342.354-.408 \cr
           & LRIS$^b$ & 32x450 $g,r,i,z$ & 58454,58455 \cr
           \cr
J0952$-$0607 
            & LRIS & 19x310 $g$, 12x300 $I$ &58455.501-.576\cr
            & & 5x300 $R$ & \cr
            & LRIS$^b$ & 12x900 $g,r,i,z$ &58454.526-.639 \cr
            & LRIS$^b$ & 4x900 $g,r,i,z$ &58577.392-.426 \cr
            \cr
J1124$-$3653&GMOS & 3x300 $r$, 2x300 $i$ & 57417.197-.228\cr
            & & 2x300 $r$, 1x300 $i$ & 57420.260-.290\cr
            & & 8x600 $g$, 8x300 $r$, 7x300 $i$& 57429.213-.323 \cr
            & & 3x423 $r$ & 56727.365-.377\cr
            & & 11x300 $i$ & 57870.151-.241\cr
            \cr
J1301+0833   & LRIS & 2x300 $I$ & 57515.490-.509\cr
             & LRIS$^b$ & 21x600 $g,r,i,z$ & 57427-57572 \cr
             & GMOS & 3x420 $r$ & 56793.969-.985\cr
             & SDSS & 1x54 $g,r,i$ & 55730.391 \cr 
             \cr
J1959+2048  & R07  & $B,V,R,I,K$  & see R07\cr
             \cr
J2052+1219  &LRIS &2x300 $g/I$  & 57305.264-.305 \cr
            & & 3x300 $g$, 4x300 $I$ & 57640.346-.366 \cr
            & & 2x300 $u/R_s$ & 57640.346-.366 \cr
            & & 6x320 $g$,7x300 $I$ & 57991.46-.99 \cr
            &DEIMOS & 2x300 $R$, 3x300 $I$ & 57686.236-.249 \cr
            &MDM &23x300 $r$ + 2x900 $r$ & 57545.343-.469 \cr
            &SAMI &  30x180 $i$   & 58336.73-.80\cr
            \cr
J2241$-$5236 &LCO 2\,m  & 9x600 $g,r,i$  &58310.127-4.260 \cr
             &SOI & 6x20, 14x180 $i$ & 57307.596-.603 \cr
             &GHTS & 3x180 $g,r,i$,  & 56514.728-5.851 \cr 
             &  & 4x300 $r$, 1x300 i& \cr
             &GMOS & 3x600 $r$, 3x300 $i$ & 58365.740-.769 \cr
\enddata
\tablenotetext{a}{Remeasured archival data.}
\tablenotetext{b}{Spectral synthesized magnitudes; see text.}
\end{deluxetable}

Observations were obtained with a variety of telescope/camera systems. In Table 2 we summarize the data used in this paper. Starting from small aperture, on the LCO 2\,m telescopes we observed with the Sinistro cameras, obtaining $gri$  measurements of our brighter sources near optical maximum. Unfortunately, these were near the system limiting magnitude; hence, low transparency, poor seeing, and tracking problems severely limited the utility of these data. In practice LCO observations only contributed to the model fitting for J0023 and J2241 (Table 2), although other measurements were broadly consistent with more-accurate photometry from other facilities. $R$-band photometry was obtained with the Tempelton CCD camera on the MDM 2.4\,m Hiltner telescope for J1301 on 2013 April 4 through 2014 May 28 (UT dates are used throughout this paper), as described by \citet{lht14}. J2052 was observed with OSMOS in the imaging mode and the SDSS $r$ filter on 2016 June 6. 

SOAR observations in 2017 were carried out with the Goodman High Throughput Spectrograph (GHTS) in imaging mode. In 2018, we observed with the SOAR Adaptive Module Imager (SAMI) which includes a UV laser system for correction of the ground-layer seeing (GLAO). A substantial portion of the run had poor seeing, where this GLAO system was of limited use, but several periods of subarcsecond seeing allow the corrections to produce a full width at half-maximum intensity (FWHM) as small as $0.5^{\prime\prime}$. Gemini GMOS data were obtained both from the science archive (for J0023, J0636, and J1301) and from new observations (under programs GS-2015B-FT3, GS-2017A-FT10, and GS-2018B-FT102).

Keck~I/LRIS observations allowed us to get simultaneous red and blue images (largely $g/I$, but some $V/I$ and $u/R_s$). A few Keck~II/DEIMOS $R$ exposures were obtained at minimum light during other programs. We also obtained Keck~I/LRIS long-slit spectroscopy of a number of BWs, during which we adjusted the slit position angle (PA) to place a nearby comparison star on the slit. After extraction and calibration of these spectra using an automated pipeline \citep{perl19}, we can integrate across broadband filters using the IRAF {\tt sbands} script to obtain simultaneous relative magnitudes for the comparison star and the BW companion. These differential magnitudes can be calibrated with the (constant) catalog magnitudes of the comparison star, to correct for variable seeing and slit losses. (However, if the slit PA is not perfectly correct, there will be a grey shift, since the slit losses will be different for the two targets.) This gives simultaneous broadband magnitudes, even when the spectral signal-to-noise ratio (S/N) was quite low. 

We obtained such spectral-synthesis magnitudes in the $griz$ bands for J0636 ($\Delta m=+0.2$\,mag), J1301 ($\Delta m = 0.1$ to $0.5$\,mag) and J0952 ($\Delta m = 0.65,\, -0.55$\,mag). The $\Delta m$ grey offsets above were determined for each epoch and applied to the resulting BW magnitudes. We estimated uncertainties from the scatter in the resulting magnitudes near maximum brightness: J0636 ($g$, 0.06; $r$, 0.03; $i$, 0.05; $z$, 0.06), J0952 ($g$, 0.05; $r$, 0.02; $i$, 0.03; $z$, 0.06), and J1301 ($g$, 0.1; $r$, 0.04; $i$, 0.03; $z$, 0.06 at maximum brightness, doubling near quadrature). The epoch-dependent grey-offset estimates above are also uncertain by as much as $\Delta m \approx 0.1$\,mag.  These spectral extractions are flagged in Table 2. At the brighter phases the exposures can have sufficient S/N for true spectral studies; these will be reported elsewhere.

In general, after standard CCD processing (bias subtraction and flat fielding),  we used IRAF to measure the FWHM of a number of bright stars to determine image quality. Then, identifying a grid of $\sim 15$--50 well-detected stars in each frame, we registered the frames to subpixel accuracy. After measuring the position of the pulsar companion in several frames near optical maximum where it was best detected, we performed forced point-spread-function (PSF) weighted photometry at the (fixed) position of the pulsar and our bright stars. The Gaussian function for the PSF weights was set to width $1.3\times$ FWHM for each frame and flux was extracted from a $1.7\times$ FWHM aperture. We then determined the catalog magnitudes of the grid stars in the relevant filter, using SDSS magnitudes directly when possible. When not, we used PanSTARRS $g,r,i,z$ magnitudes, converting to SDSS using the transformations recommended by \citet{fet16}:
$$
m_{\rm SDSS} = m_{\rm P1} - (a_0 +a_1 x + a_2 x^2 +a_3 x^3), \qquad x=g_{\rm P1}-i_{\rm P1},
$$
where the $a_i$ coefficient vectors are as follows: $u$ (0.04438, $-2.26095$, $-0.13387$, 0.27099) [correcting from $g_{\rm P1}$], $g$ ($-0.01808$, $-0.13595$, 0.01941, $-0.00183$), $r$ ($-0.01836$, $-0.03577$, 0.02612, $-0.00558$), $i$ (0.01170, $-0.00400$, 0.00066, $-0.00058$), and $z$ (0.08924, $-0.20878$, 0.10360, $-0.02441$). For two of our southern pulsars (J1124 and J2241) we have neither SDSS nor PanSTARRS coverage, and so we were forced to use APASS photometry to transfer from bright stars to our infield star grid. We obtained $g,r,i$ frames with the LCO 1\,m system, where the large field provided more APASS stars for calibration of the nearby fainter comparison star grid.

These BWs have magnitude ranges as large as $\delta m = 5$\,mag Table 1). At maximum brightness, the photometric statistical errors can be unrealistically small, $<0.01$\,mag. Therefore, we add a systematic $\sigma_s = 0.03$\,mag in quadrature to all uncertainty estimates (except for those of the brightest BW, J1959) before the light-curve fitting to account for unmodeled zero-point, flat-fielding, and aperture errors. Conversely, at minimum brightness some objects, while formally detected in our forced (fixed position) astrometry, have very large statistical uncertainties ($\sigma_m \approx 1$\,mag). While we do include these points in our model fits, all detections with $\sigma_m > 0.5$\,mag should probably be properly viewed as upper limits on the companion flux.

After barycentering the time of all observation midpoints and computing the orbital phases with the latest ephemerides available, we plot the folded light curves in Figures 3--5, with two cycles shown. Phase 0 is at the pulsar ascending node, so that pulsar inferior conjunction (near optical maximum) is at $\phi_B=0.75$.

\begin{figure*}[t!!]
\vskip -3mm
\includegraphics[width=9cm]{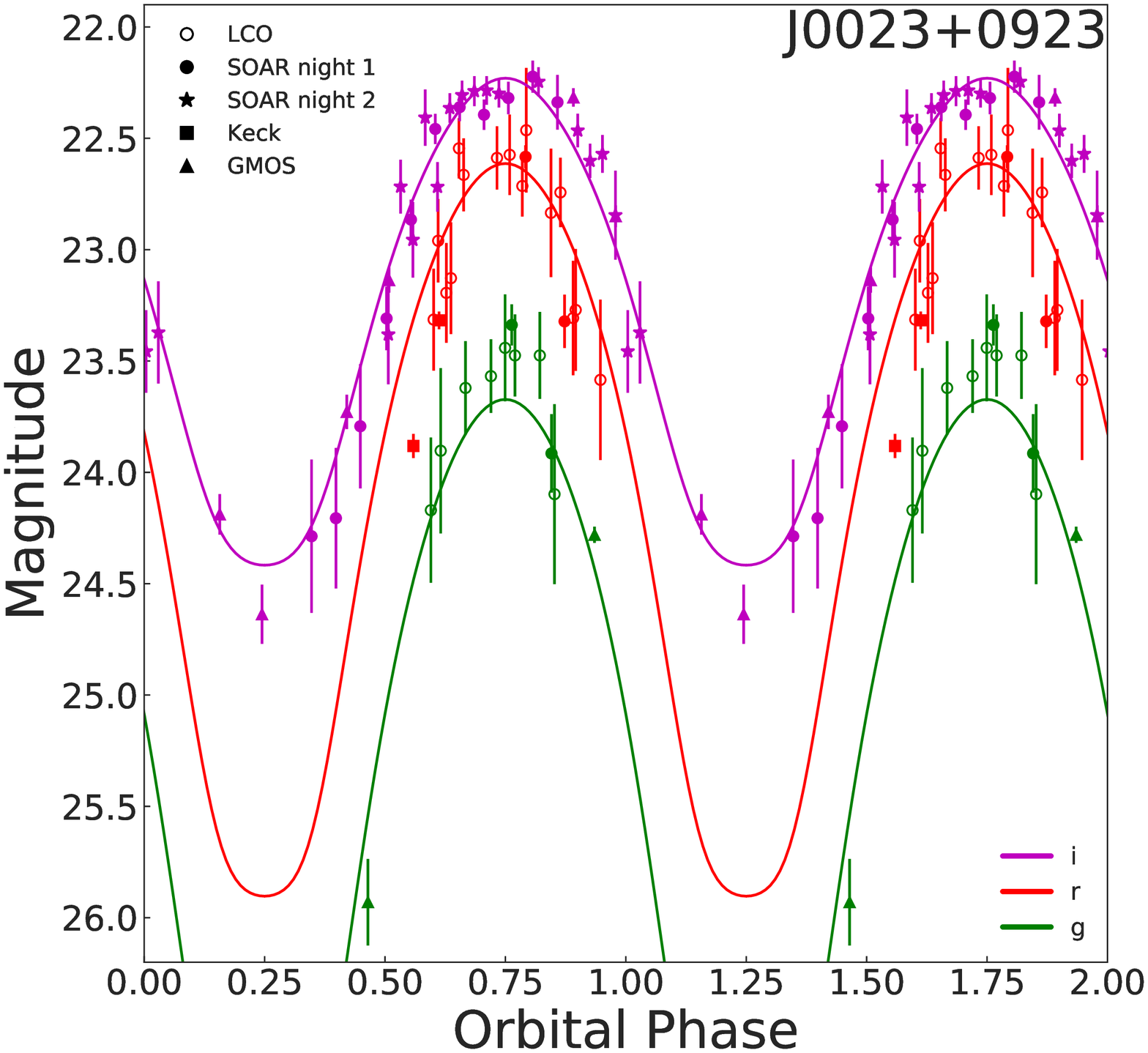}\includegraphics[width=9cm]{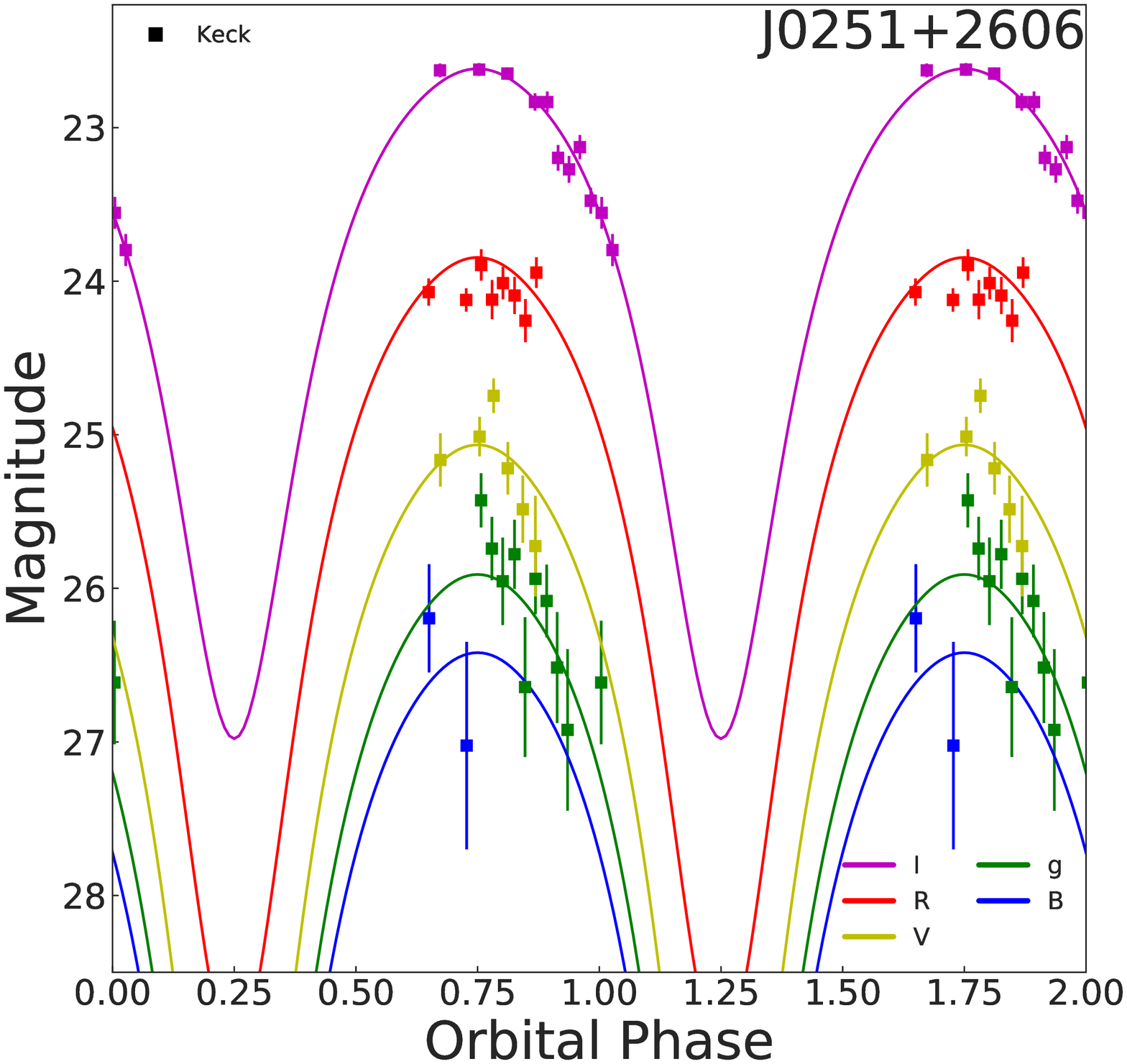}
\vskip -4mm
\includegraphics[width=9cm]{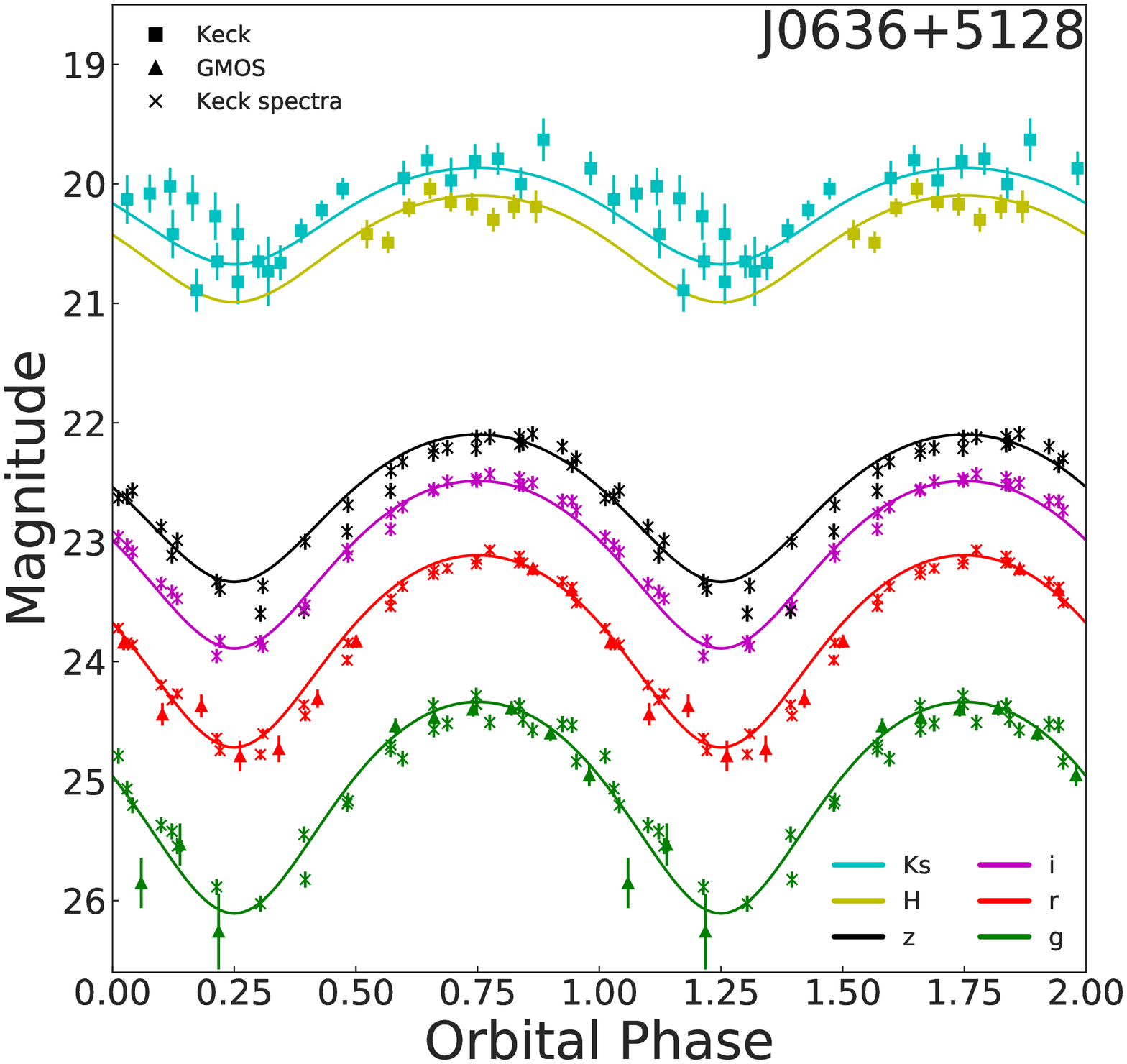}\includegraphics[width=9cm]{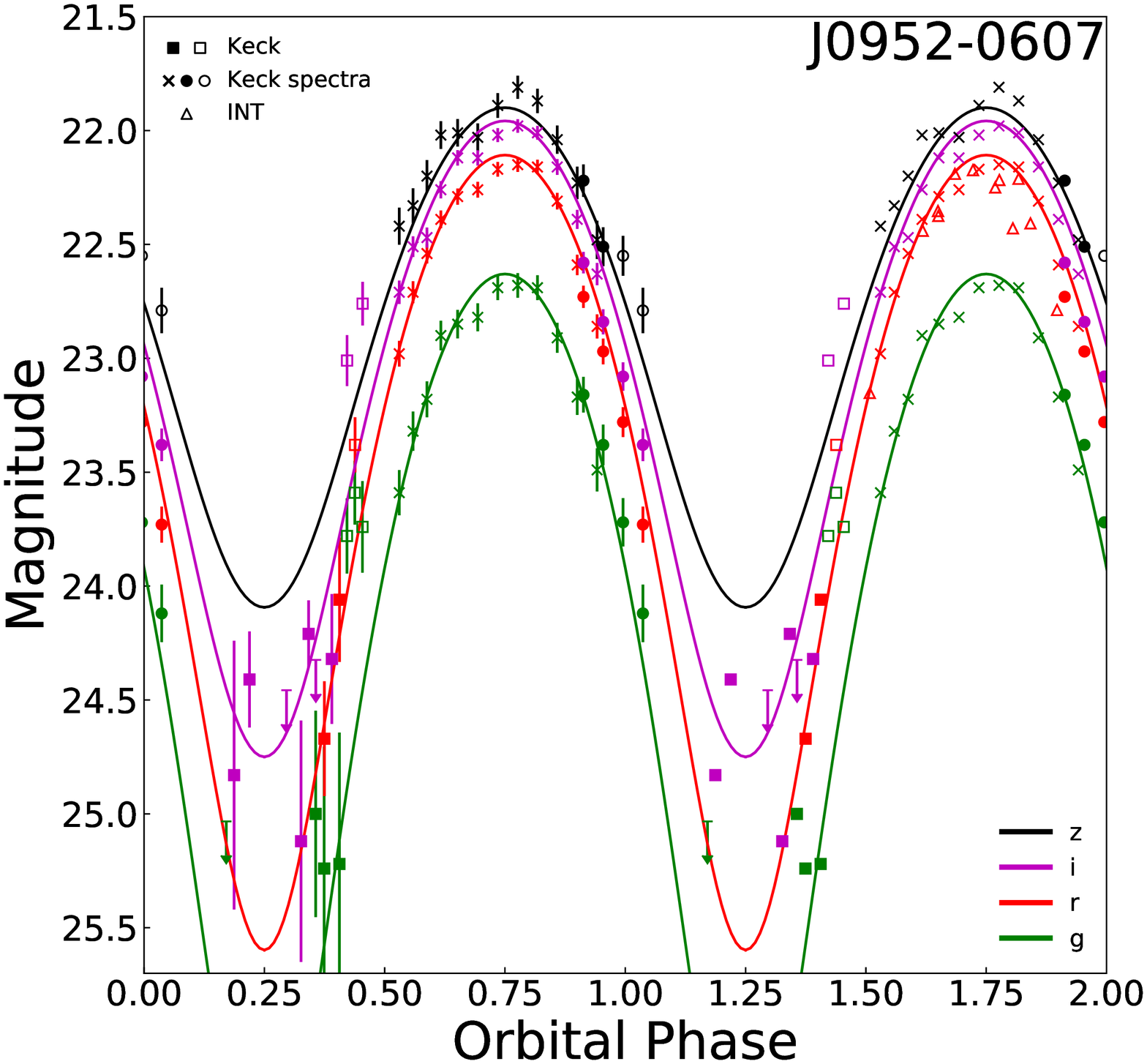}
\caption{\label{LCs1} Four BW light curves. Two periods are plotted; $\phi_B=0$ is at pulsar TASC (time of the pulsar ascending node). For J0952, photometry points during the ``flare'' at $\phi_B=0.4$--0.5 are not used in the fit. These are marked with open symbols. During the second cycle the error flags are omitted and the INT $r$ photometry is shown, for comparison.} 
\end{figure*}

\begin{figure*}[t!!]
\vskip -3mm
\includegraphics[width=9cm]{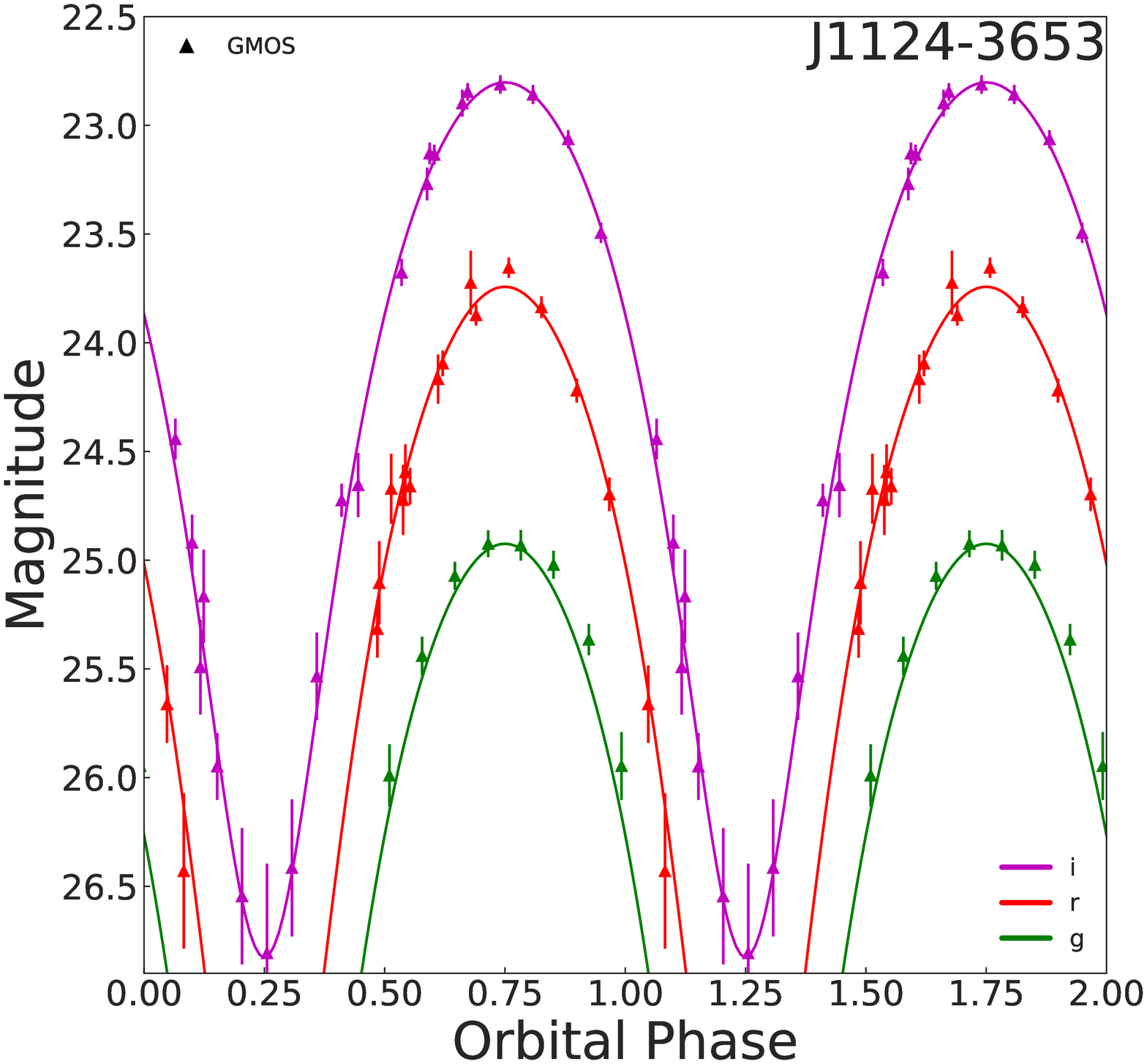}\includegraphics[width=9cm]{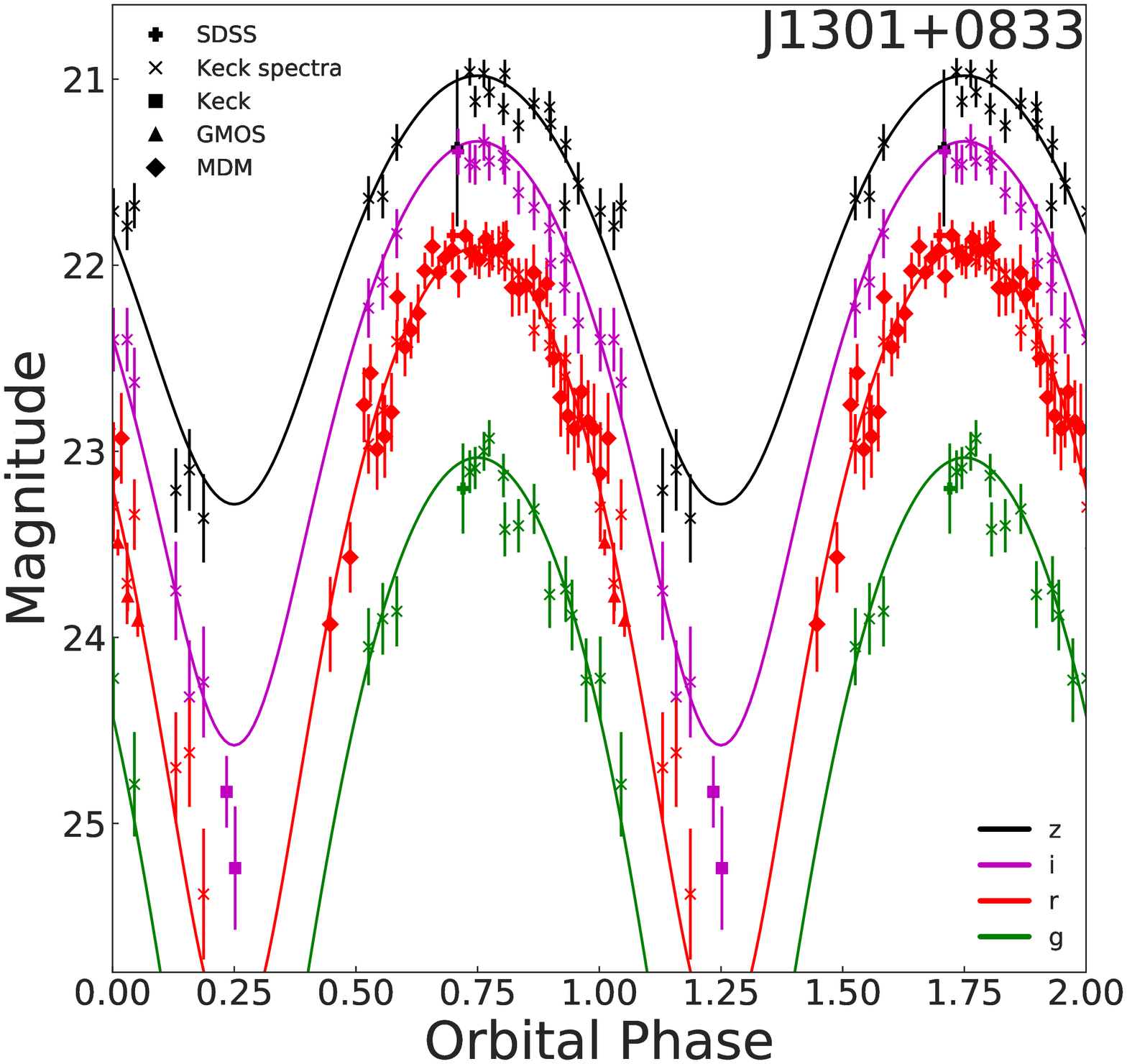}
\vskip -4mm
\includegraphics[width=9cm]{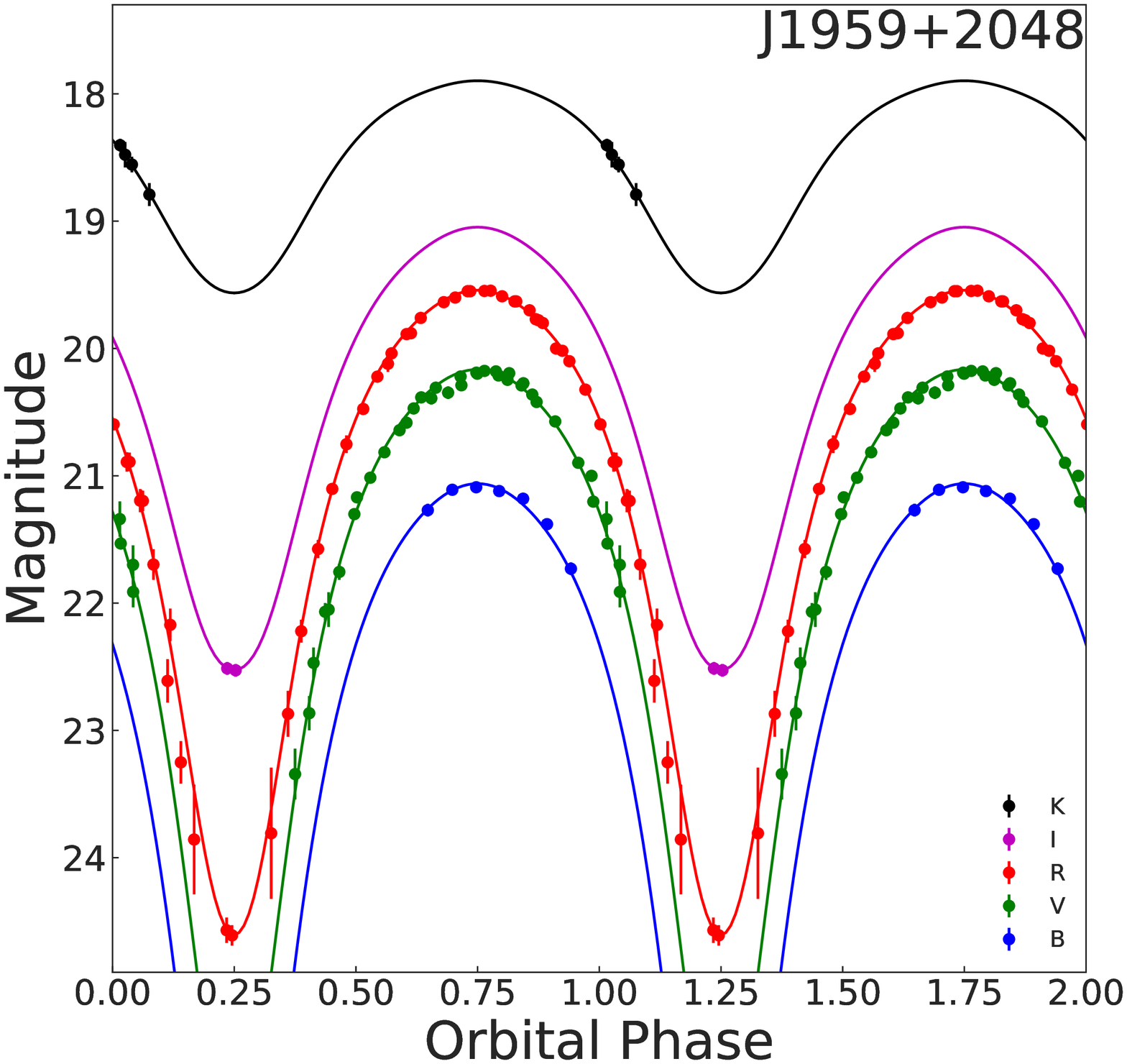}\includegraphics[width=9cm]{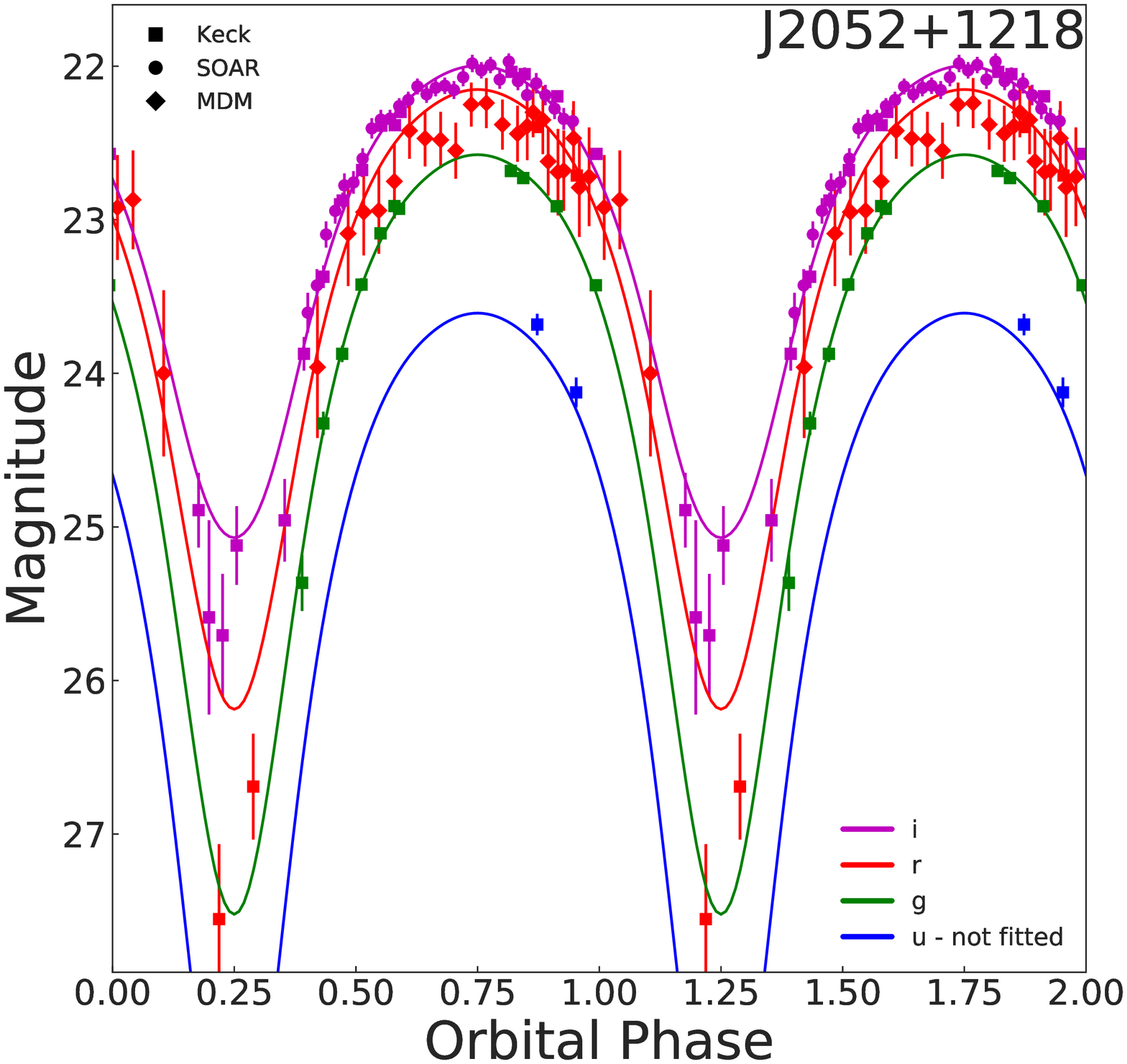}
\caption{\label{LCs2} Four BW light curves. Two periods plotted; $\phi_B=0$ is at pulsar TASC (ascending node). J2052 MDM $r$ statistical errors have been doubled.} 
\end{figure*}

\begin{figure}[h!!]
\vskip -3mm
\includegraphics[width=9cm]{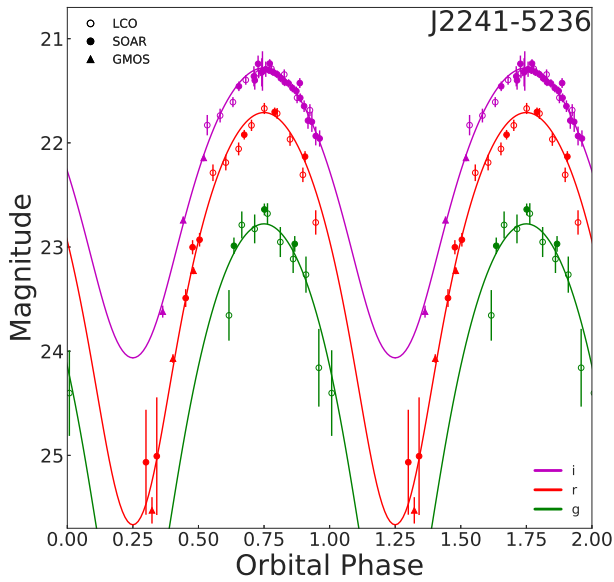}
\caption{\label {LC3}J2241 light curve.} 
\end{figure}

\begin{figure*}[t!!]
\includegraphics[width=9cm]{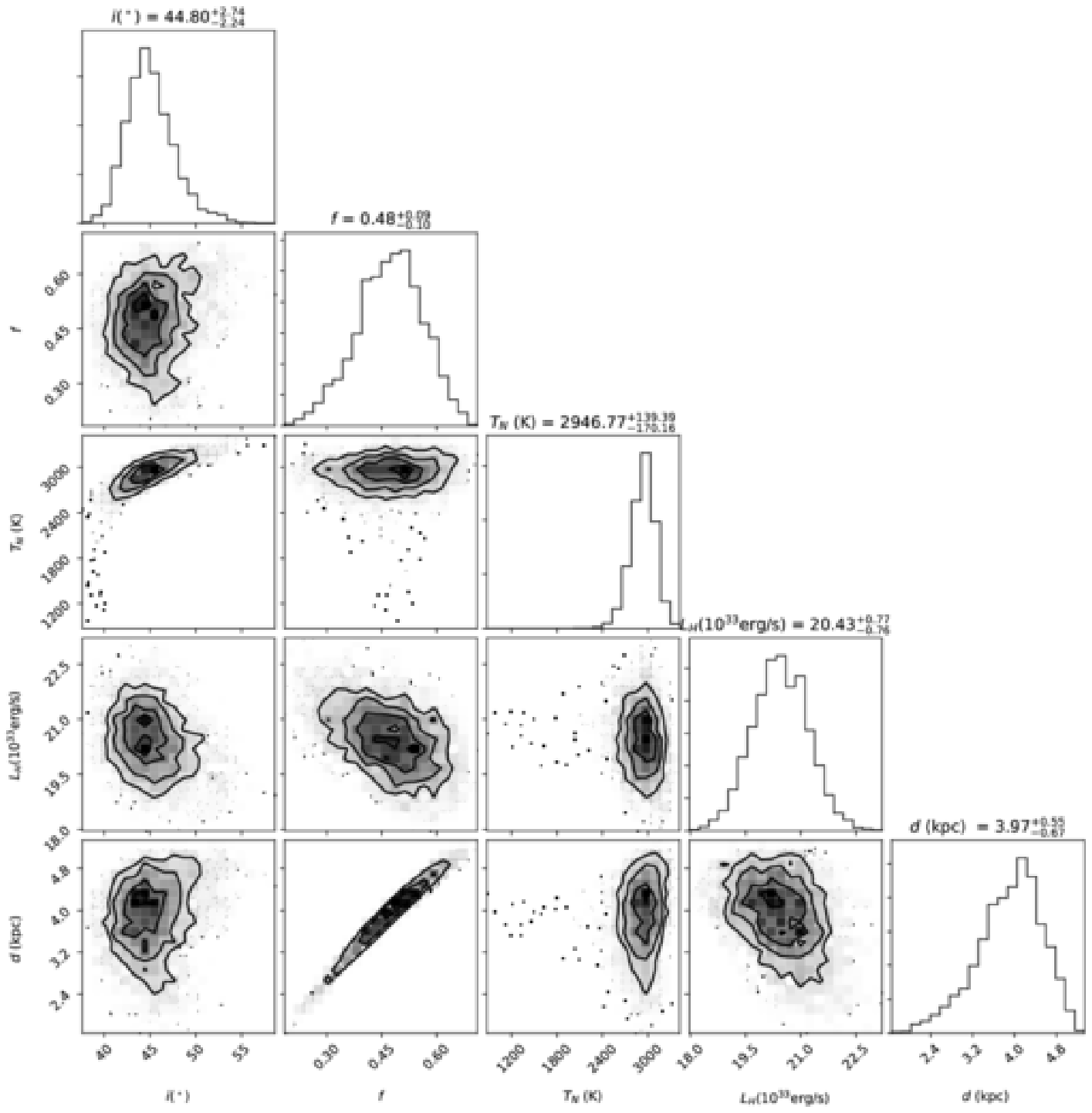}
\includegraphics[width=9cm]{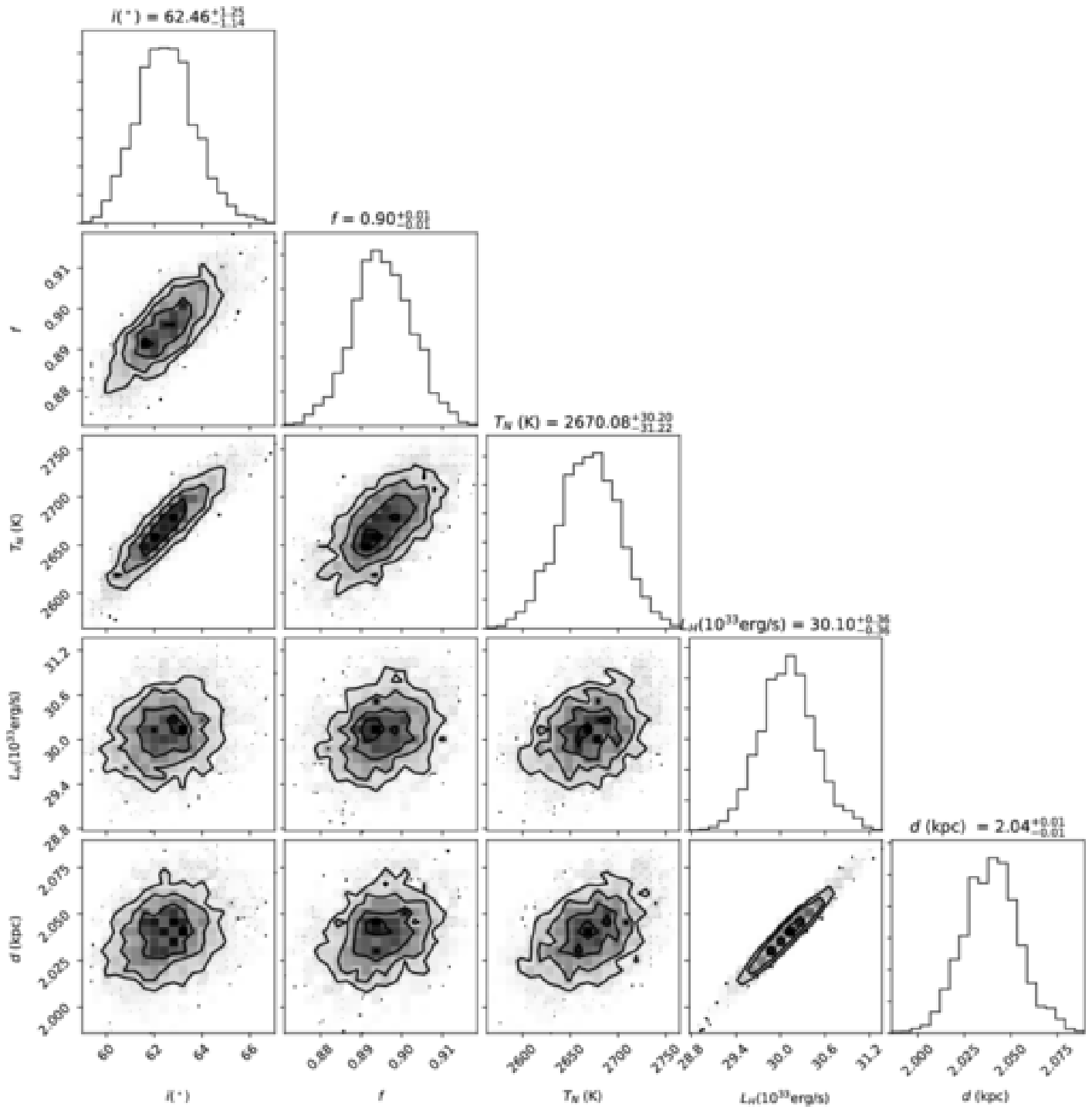}
\caption{\label {MCMC} Left: corner plot for J0952. Inclination $i$ in degrees, $L_X$ in $10^{33}\,{\rm erg\,s^{-1}}$, and distance in kpc. Note the strong $f_1$--distance correlation. Right: corner plot for J1959, with well defined maxima.} 
\end{figure*}

\section{Direct-Heating ICARUS Fits}

We fit these data with the ICARUS modeling code \citep{bet12} in its original form, where a bolometric luminosity $L_X$ is assumed to propagate linearly from the pulsar point source and heat the companion. However, most of the pulsar spindown power is emitted as a relativistic magnetized $e^\pm/B$ wind, rather than as photons. This wind is typically reprocessed in an intrabinary shock, and the resulting energetic particles can irradiate the surface or even be ducted by companion magnetic fields to surface caps. Such effects are described by \citet{rs16} and \citet{sr17}, but for simplicity and consistency we here focus on the basic direct-heating fits; such model extensions may, however, be important for more realistic fits.

The direct-heating model takes as its basic parameters $L_X$, the source distance $d$, the companion Roche-lobe fill factor $f_c$, the uniform underlying temperature of the companion $T_N$, and the binary inclination $i$. $L_X$ is reported assuming isotropic emission over the sky, but we suggest a model-dependent beaming correction for this in \S4.1; the heating effect is from the flux directed at the companion. Our fit source distance can be compared with the radio dispersion measure (DM) distance estimate from Table 1. The fill factor is the ratio of the companion radius at the nose to the $L_1$ radius. Note that for the substellar BW companions there is no nuclear energy source. However, tidal effects can provide some internal heat and conduction must carry some energy from the day side to the night side, so $T_N$ should be small but nonzero. In the model fitting, we use the precise orbital period $P_b$ and the projected pulsar semimajor axis $x_1$ from the radio and/or gamma-ray ephemeris. For J1301 we can use the published estimate for the companion radial-velocity amplitude $K=259 \times 1.08 = 280 \,{\rm km\,s^{-1}}$ \citep{rgfz16}. Similarly, for J1959, \citet{vKBK11} give $K=324 \times 1.09 =353\, {\rm km\,s^{-1}}$, where the observed velocity is corrected by a geometrical correction factor $K_{\rm cor}\approx 1.1$ to account for the difference between the radial amplitude of the spectrally-weighted center-of-light motion and the desired center-of-mass motion. With this we get an estimate of the mass ratio $q=KP/(2\pi x_1)$. 

For our other BWs, no radial velocity is yet published; we must assume an NS mass. \citet{antet16} found that the MSP distribution is bimodal with a dominant peak at $M_{\rm NS} \approx 1.4\, M_\odot$ and a subpopulation with $M_{\rm NS} \approx 1.8\, M_\odot$ (width $\sigma_{M_{\rm NS}} \approx 0.2\, M_\odot$). \citet{stret19} find that essentially all RB MSPs were from the higher mass peak. Here we assume a round  $M_{\rm NS} = 1.5\, M_\odot$ to acknowledge that evolution and accretion probably increase the BW masses above the normal pulsar $1.4\,M_\odot$, without tying to a specific value. If one assumes that BWs, like RBs, populate the higher mass peak, our companion masses would increase by $\sim 13$\%. A more serious estimate of the pulsar mass distribution should be attempted once more BW radial velocities are available. This requires careful treatment of the radial velocity correction factor, so we do not discuss it further here.

The observed colors are affected by the intervening extinction $A_V$. For many of our objects, we can adopt the extinction estimates from the three-dimensional dust model of \citet{gsfet15} available at \text{argonaut.skymaps.org}, using the Bayestar2019 model and $A_V=2.73\times E(g-r)$ values at the estimated distance. For the two southern sources, we can use as an upper limit the \citet{sf11} total extinction column through the Galaxy from NED. 

To obtain observed-band fluxes from the companion temperature distribution, we use a set of model-atmosphere spectra folded through the responses of our broadband filters, as provided by the Spanish Virtual Observatory (svo2.cab.inta-csis.es). As in the past, we use the BTSettl atmospheres of \cite{allet12}, to cover the very large temperature range from the heated day side to the cool night side, assembling a table of $F_k(T,g)$ surface fluxes in each band $k$. Since spectroscopic studies of BW (and RB) companions indicate that the stripping can leave remnants with appreciable metal enrichment, we also explore the metallicity dependence of the model fits, using the available log\,$[Z] = -1$, $-0.5$, 0, +0.3, and +0.5 BTSettl models. Finally, we identified a flux-normalization error in our previous version of the code that resulted in $d$ estimates being too large in \citet{sr17}; this has been amended and the direct-heating distances derived here should supersede those values.

As noted above, we add $\sigma_s =0.03$\,mag in quadrature to each point before fitting (except for J1959), to reduce sensitivity to the very small statistical photometric uncertainties at maximum brightness. We start the fitting with the python scipy least-squares minimization and then map the minimum and covariances with Markov Chain Monte Carlo (MCMC) runs (MultiNest v3.10). Characteristic corner plots from the latter are shown in Figure \ref{MCMC}. The model filter light curves for the resulting best fits are shown in Figures 3--5 while the fit parameters are listed in Table 3.

The overall matches to the light curves are reasonable, with strongly heated companions brightest near $\phi_B \approx 0.75$, the subpulsar points. However, the $\chi^2/{\rm DoF}$ are relatively large for many of these fits. This may indicate that the photometric uncertainties are underestimated or that there is random variation about the quiescent light curve \citep[e.g., flaring as seen for PSR J1311$-$3430;][]{ret15}. It is also likely that there are physical effects in the companion heating not captured in the basic model.

\begin{figure}[h!!]
\includegraphics[width=9cm]{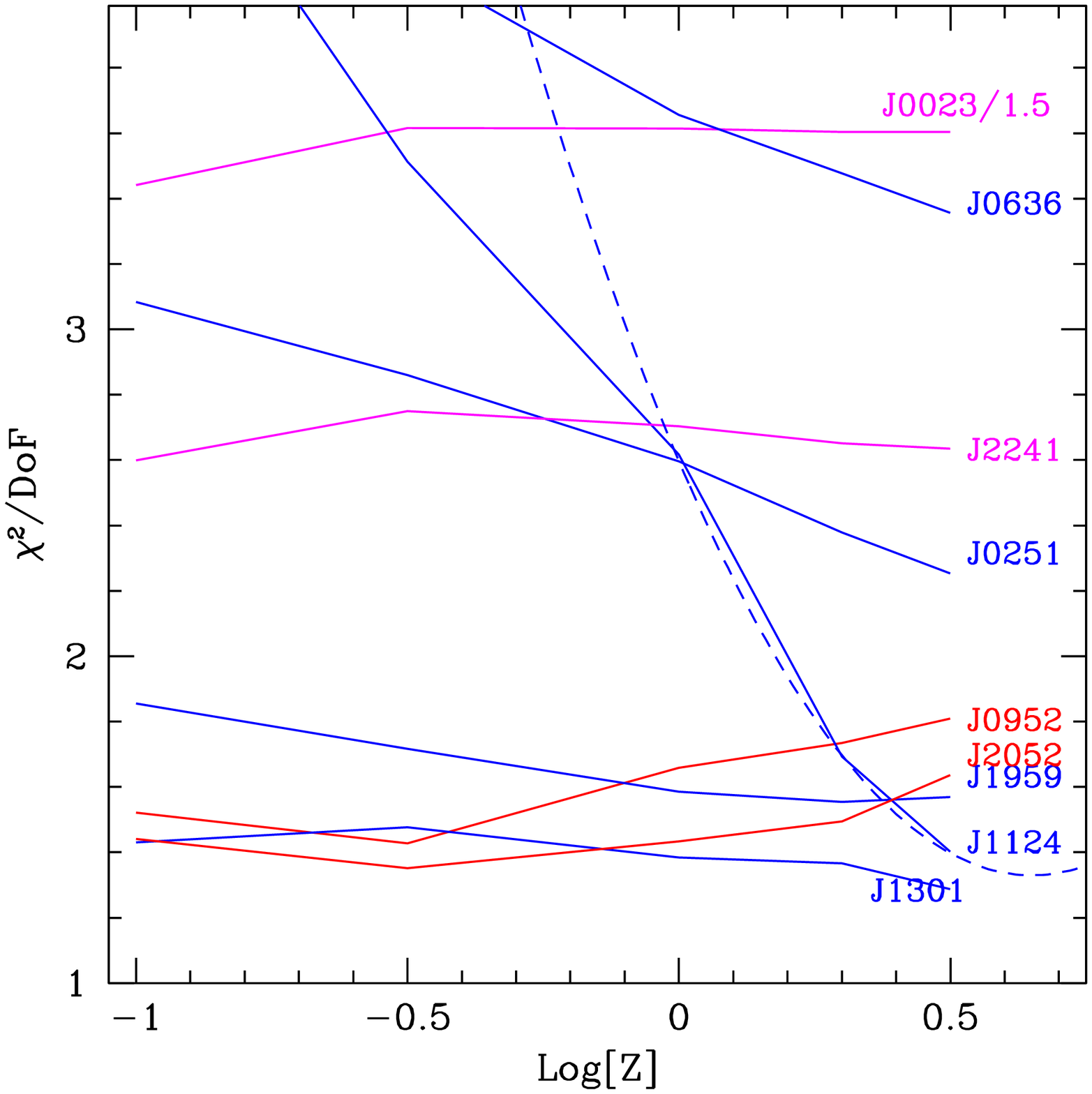}
\caption{Metallicity dependence of the model fits. Higher metallicity is generally preferred (blue lines), although two BWs seem to prefer lower log\,$[Z]$ (red lines). For the strongest dependence, J1124, we can extrapolate from the last few points to infer a minimum at log\,$[Z] \approx 0.65$. For J0023, we plot $\chi^2/1.5\times$DoF. }
\end{figure}

Five of our BWs (J0251, J0636, J1124, J1301 and J1959) have systematically decreased $\chi^2/{\rm DoF}$ using atmosphere tables with higher metallicities. Two (J0952 and J2052) appear to prefer log\,$[Z]\approx 0$ or lower. Interestingly, these two BWs have hottest day-side temperatures in our sample, so it may be that in this higher $T$ range the colors are less sensitive to metal overabundance. The last two objects show no significant dependence. The models only extend to log\,$[Z]=+0.5$, or $\sim 3\times$ solar, but for the strongest dependence, in J1124, we can extrapolate from the most metal-rich models to infer a minimum at $\sim 4.5\times$ solar abundance, although all one can really conclude is that log\,$[Z]>0.5$ should produce even better fits for several of our pulsars (Figure 7). 

\subsection{Spindown Power and Beaming Corrections}

Because we assume that the companion is directly (radiatively) heated by the pulsar, it is useful to compare the results of the photometric heating model with other measures of the pulsar output. In fact, the $\nu F_\nu$ spectral energy distributions (SEDs) of rotation-powered pulsars are strongly dominated by the GeV peak observed by the {\it Fermi} LAT, so the observed LAT fluxes (Table 1) provide an estimate of the incident photon flux (scaled with the distance). Heuristically, the LAT GeV luminosity scales with pulsar spindown power ${\dot E}$ as $L_{\rm heu} \approx (10^{33}\,{\rm erg\,s^{-1}} {\dot E})^{1/2}$ (independent of binary properties). This means that the GeV efficiency $\eta_{\rm heu} \approx (10^{33}{\rm erg\,s^{-1}}/{\dot E})^{1/2}$ is $\sim 0.01$--0.3 for most $\gamma$-ray pulsars.

The spindown power should be corrected for the Shklovskii effect,
$$
{\dot E}_c=(10^{34}\,{\rm erg \,s^{-1}}){\dot E_{\rm 34}} I_{45} [1-0.0096\mu_{\rm mas/yr}^2 d_{\rm kpc}/({\dot E_{\rm 34}}P_{\rm ms}^2)],
$$
where ${\dot E}_{34}$ is the nominal spindown luminosity in units of $10^{34}\,{\rm erg\,s^{-1}}$ for a moment of inertia $10^{45}I_{45}\, {\rm g\,cm^2}$, $\mu_{\rm mas/yr}$ is the total proper motion in mas\,yr$^{-1}$, $P_{\rm ms}$ is the spin period in ms (from Table 1), and $d_{\rm kpc}$ is the fit (or DM) distance. This correction can be quite substantial; for example, at the nominal DM distance J2052's spindown power would be reduced by a factor of 3, decreasing the energy budget available for companion heating.

Since the model fits estimate this heating luminosity, it is useful to compare with the heuristic expectation $L_{\rm heu}$ and the observed gamma-ray power. However, the LAT gamma-ray flux is that directed at Earth's line of sight at inclination $i$, while the heating flux is directed at the companion star, which for spin-aligned MSPs will be at the pulsar equatorial plane. Since in most models pulsar emission is not isotropic, we must correct these fluxes to isotropic all-sky equivalents with ``beaming corrections'' $f_i$ and $f_\ast$, respectively, to make such comparisons.

We expect the pulsar magnetosphere flux to be directed toward the spin equator, as seen in numerical models \citep[e.g.,][]{tchet16}. There the energy/momentum flux of the pulsar wind is concentrated as $\propto \sin^n\theta_*$ with $n=2$ or even $n=4$ for an oblique rotator ($\theta_*$ is the angle from the pulsar spin axis, assumed perpendicular to the orbital plane). This flux is largely particles and fields, but the direct (photon) radiation should be similarly concentrated. Gamma-ray pulsar models typically have the emission occur near the light cylinder or even in the near-field wave zone outside the light cylinder \citep[e.g.,][]{kalet18}. In such models the radiation beamed from these active zones is also thought to be equatorially concentrated. Note that there may also be (evidently fainter) emission beamed along the dipole magnetic axis, since lower-altitude radio pulses imply pair production (and hence beamed gamma-ray emission) in this zone, as well.

\begin{figure}[t!!]
\hskip -4mm
\includegraphics[width=4.6cm]{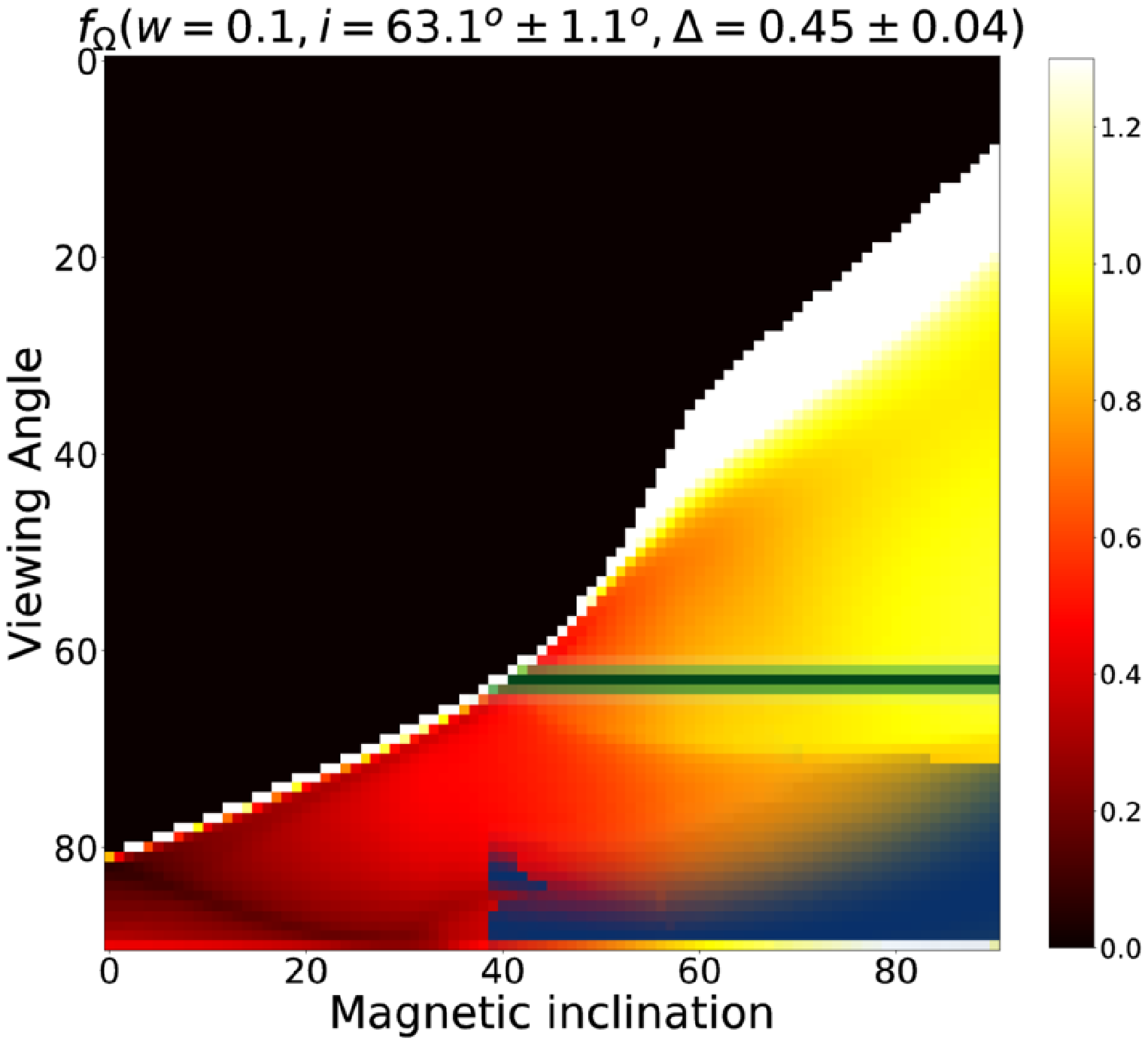}
\hskip -4mm
\includegraphics[width=4.6cm]{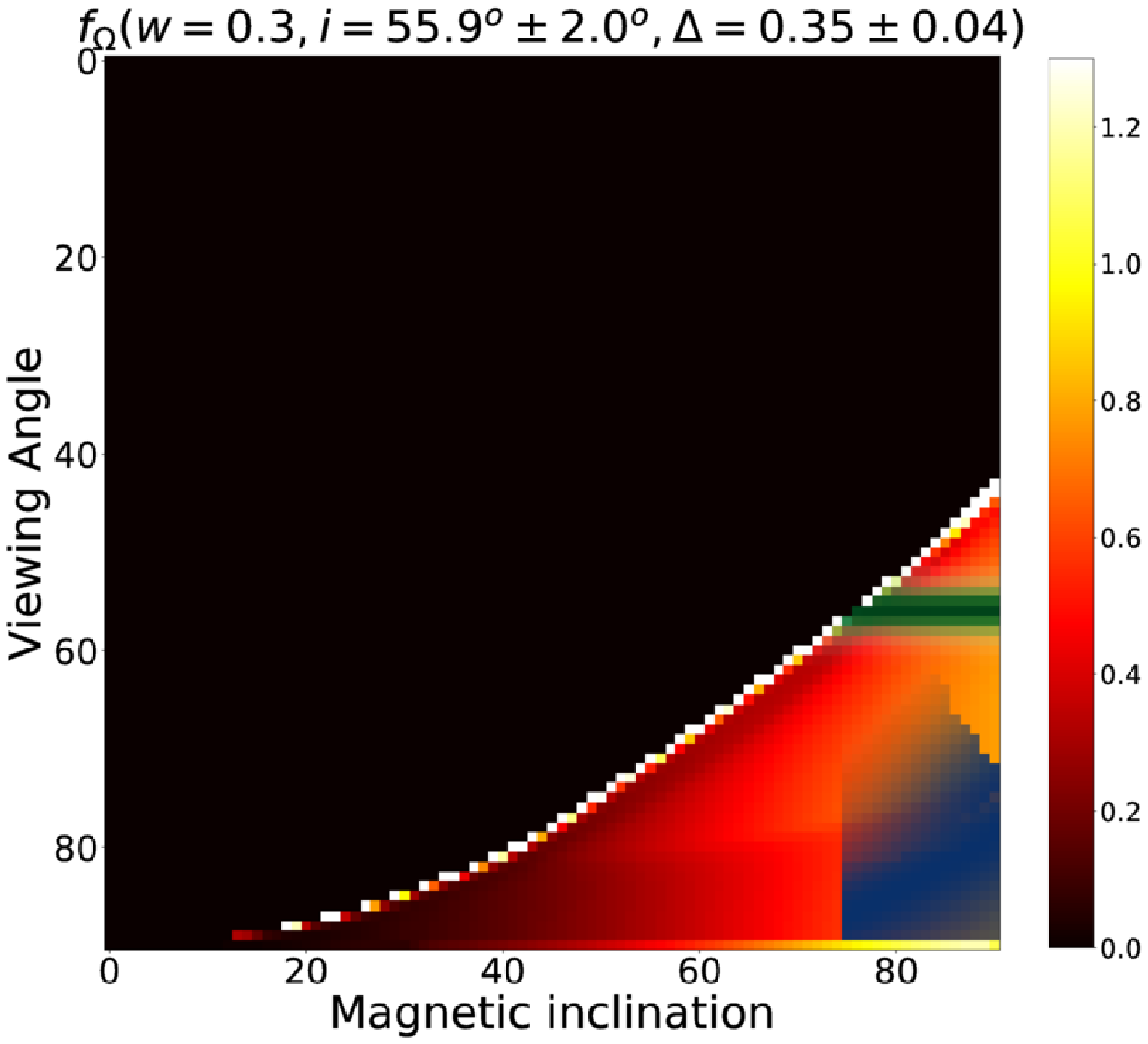}
\vskip -4mm
\caption{\label{fOm} Flux correction factor $f_\Omega (\alpha, i)$ for two BWs (left, J1959; right, J2052), where the abscissa maps the magnetic inclination $\alpha$ and the ordinate is the viewing angle $i$ (color map of $f_\Omega$ values to the right of each frame). The inclination range $i$ from our photometry fits is shown by the green band (truncated to the $\alpha$ that produce a visible $\gamma$-ray pulse); the blue overlay shows the regions best compatible with the observed gamma-ray pulse separation $\Delta_\gamma$ for an outer gap model. Note that higher $i$ are more natural in this model and that the observed $\Delta_\gamma$ does not occur for $i$ as small as found in the photometric fits.} 
\end{figure}

For a concrete example, we use the geometrical ``outer gap'' (OG) model whose beaming factors $f_\Omega(\alpha,\zeta, \eta_{\rm heu}) = f/f_{\rm iso}$ are given by Figure 16 of \citet{rw10} as a function of magnetic inclination $\alpha$, viewing angle $\zeta=i$, and heuristic $\gamma$-ray efficiency $\eta_{\rm heu}$. These are meant to correct observed fluxes to full-sky isotropic emission equivalences, but they also provide the model-dependant correction for the flux viewed at $i$ to the flux directed at the companion near $\zeta = \pi/2$ (see Table 3, right section). Our fit $i$ has an uncertainty range (and the companion subtends nonzero angle), so the correction $f_i$ is computed by weighting over the $i$ uncertainty range and scanning over the $\alpha$ range which (for that $i$) allows Earth to see the $\gamma$-ray beam. For the companion heating we compute $f_\ast$ by integrating over the angle subtended by the companion, and scanning the allowed $\alpha$. We assume that angles are distributed isotropically within the allowed ranges. Two examples are shown in Figure \ref{fOm} and the values are listed in Table 3. Note that these values are averaged over a range of angles and so, understandably, are rather close to 1; true values for a specific pulsar's $\alpha$ and $i$ can be quite far from unity.

We find it convenient to divide the (beaming-corrected) gamma-ray flux and heating flux by the (Shklovskii-corrected) spindown power and thus refer to the gamma-ray ``efficiency'' $\eta_\gamma = f_\gamma f_i 4\pi d^2/{\dot E}_c$ for the sky output of $\gamma$-rays and $\eta_H = L_x f_*/{\dot E}_c$ for the output as estimated from the heating flux (see Table 3). If the measurements were perfect and the OG model for the GeV beaming were correct, these would be the same. Since $\eta_\gamma$ is an Earth-based estimate it depends on $d$, while $\eta_{\rm H}$ is measured in the BW system and so is (after Shklovskii correction) distance-independent. Inconsistency indicates bias in the light-curve fits, imperfections in the beaming model, or both. We evaluate the consistency in \S4.3.

\subsection{Individual Data and Fit Results}

{\bf PSR~J0023+0923} had a few GMOS images reported by \citet{bet13}, which we remeasured and use in our fits. We made a number of LCO 2\,m queue observations, but tracking in this region was exceptionally poor and so only a small fraction of the exposures provided usable photometry. A few Keck/DEIMOS exposures provided higher S/N detections in $R$, and two long sequences of SAMI $i$ imaging provided good coverage at maximum brightness. These $i$ data are interesting as an abrupt step occurs near maximum at $\phi_B=0.8$; this feature is consistently present on two consecutive nights in the middle of continuous image sequences. It is hard to imagine such sharp light-curve features from any thermal surface emission; one possible origin is beamed emission from an intrabinary shock. Additional observations, especially simultaneous multicolor photometry and sensitive X-ray light-curve measurements, can test whether there is a contribution from non-thermal emission at this phase. 

The few points near minimum brightness (especially the GMOS $g$ values) drive our solution to large inclination $i$. The new $i$-band magnitudes support this solution, but imply a relatively large night-side temperature. \cite{bet13} noted that this BW does not display radio eclipses and with a handful of points they find a smaller $i=58\pm 14^\circ$, which is $\sim 1.5\sigma$ away from our fit minimum, requiring $\Delta \chi^2=+60$ with our data. The J0023 fit has the worst reduced $\chi^2$ of our set, apparently due to an under-prediction of the $g$ maximum, and an indication in $g$ and $r$ that the maximum is narrower in the bluer colors. Additional photometry at minimum, perhaps in the $z$ band, would help constrain $T_N$. Higher S/N multicolor light curves are needed to map the maximum and constrain the nonthermal emission, if present.

{\bf PSR~J0251+2606} has poor light-curve coverage and, lacking detections near minimum brightness, the inclination is not well constrained. Moreover, the relatively good photometry near maximum suggests significant short-term variability, especially in the bluer $V$ and $G$ data. The $R$ and $I$ light curves are relatively well behaved. The most puzzling aspect of the fit is the large distance (2.8 times the DM estimate), and the resulting large inferred $\gamma$-ray luminosity and direct heating power, which are 47\% and 23\% of the spindown power, respectively (see \S4.3). At present we do not have a full proper motion, but Shklovskii correction using the lower limit on the right-ascension component already substantially increases the implied efficiency. The $\gamma$ and heating flux estimates would be in agreement at $d\approx 2.3$\,kpc, so a smaller distance seems likely. Clearly additional measurements, especially at minimum brightness, are needed.

{\bf PSR~J0636+5128} has a relatively shallow light curve, and both \citet{dr18} and \citet{kapet18} inferred a small inclination $i$, as also found here. An early timing parallax \citep{stoet14} gave a very small distance, but this has been superseded by more extended timing, giving $d=1.1^{+0.6}_{-0.3}$\,kpc \citep{arzet18}, and our photometric distance fit is in good agreement with this value. While not in the 4FGL source catalog (hence the upper limit in Table 1), it does have a LAT pulse detection \citep{smithet19} and is thus a (faint) GeV emitter. The required direct heating $L_X$ is only 10\% of the spindown power, but exceeds the observed $\gamma$-ray flux, as discussed below. Note that with this very small $i$, the estimated companion mass is not especially small, with $M_c \approx 0.018 (M_p/1.5\,M_\odot)^{2/3}$. If the pulsar is massive, J0636 is closer to the BW range than most Tidarrens. Spectroscopic measurement of the companion composition and radial velocity are needed to test whether it is the Tiddaren descendent of an UCXB. The large $\chi^2/{\rm DoF}$ is mostly due to a significant phase shift (data late by $\Delta \phi = 0.02$; applying this arbitrary shift reduces $\chi^2$ by $\sim 200$).

{\bf PSR~J0952$-$0607:} In \citet{baset17}, a partial $r$ Isaac Newton Telescope (INT) light curve showed the expected $\phi_B=0.75$ maximum, but placed few constraints on system parameters. We obtained (Dec. 2018) Keck/LRIS spectra across maximum brightness that provide $griz$ colors, supplemented by two-color LRIS imaging during minimum (on the next night), and a few more LRIS spectra in April 2019. These provide appreciable color data. We determined the spectral gray shifts by a match to the INT $r$ light curve, showing that the slit placement was not always optimal. The last three imaging points ($2\times gI$, $1\times gR$, marked in Figure 3) are substantially brighter than any reasonable extrapolation of the prior day's light curve --- we infer that J0952 was undergoing a flare event at $\phi_B\approx 0.4$--0.5, such as seen in other BWs. Also, relatively poor sky subtraction beyond 8500\,\AA, affected the $z$-band estimates of the last two spectral measurements, so we exclude these points. As this paper was being submitted for publication, \cite{nidet19} presented observations of J0952 that include ULTRACAM/HiPERCAM photometry. Although presented in flux rather than magnitudes, it appears that these data also include a similar (red) flaring event, in a similar phase range.

J0952's peak appears relatively narrow; thus, our fits prefer small $f_c$ with minimal peak-broadening due to ellipsoidal modulation. This is the only object for which the fits prefer $f_c<0.5$; for such small values the companion is nearly spherical and the distance is highly covariant with $f_c$, with little constraint on either. The very best fit occurs for log$[Z]=-0.5$, at which the solution drives to unphysically small $d$ and $f_c$. This appears to be due a $z$-band offset of $\sim -0.02$\,mag for this metallicity, which better matches the data. For both smaller and larger log$[Z]$ the fits give $d \approx 3$--4\,kpc, but the $d$ minima are shallow, allowing smaller distances down to the DM estimate of 1.7\,kpc with $\Delta \chi^2\approx+8$. However, larger $d$ are relatively strongly excluded; with the $5.6$\,kpc of \cite{nidet19} we find $\Delta \chi^2=+50$. We conclude that the very small $d$ of the log$[Z]=-0.5$ fit may be an artifact of imperfect photometric calibration and adopt here the agnostic (but slightly worse) log$[Z]=0$ fits. Future precision photometry, especially in the near-infrared (or a parallax measurement), will be needed to study this important object. All parameters other than $d$ and $f_c$ were nearly independent of metallicity, so we consider these to be relatively robust.

{\bf PSR~J1124$-$3653} was observed in several visits through the Gemini Fast Turnaround Queue program. One observational challenge is the very bright $R=12$\,mag star at the same declination, whose saturated core caused a charge-transfer-efficiency (CTE) trail to pass very close to the pulsar in initial observations (Figure 2). For exposures capturing minimum brightness, we placed this star behind the guider pick-off mirror to avoid this issue. Our ICARUS fit is quite good, although the $g$ data peak appears somewhat broader than that of the model. The main challenge is a fit distance $\sim 2.5$ times larger than the DM estimate. Since the spindown luminosity is small (even before Shklovskii correction), the required $\gamma$-ray heating power is $2.4\times$ the nominal spindown luminosity. The $\gamma$-ray flux and direct-heating power at $d=1.2$\,kpc are in better agreement with the DM estimate. However, a direct-heating fit with $d$ fixed at this value is much worse, with $\chi^2/{\rm DoF} \approx 2.5$ times larger, and $i$ decreased in an (unsuccessful) attempt to fit the broad maximum.  We suspect that a departure from simple direct heating is needed to accommodate a smaller distance. 

{\bf PSR~J1301+0833}'s light curve has points converted to $r$ from \citet{lht14}, but the rest are the imaging and synthesized magnitudes discussed by \citet{rgfz16}. Here our fit is not very sensitive to log\,$[Z]$, but we do achieve a lower $\chi^2$ and a closer distance estimate than found in the latter paper (although still larger than the DM estimate). At this 2\,kpc distance the corrected spindown power is ${\dot E}_c=2.5 \times 10^{34}\,I_{45}\, {\rm erg\,s^{-1}}$, so the heating flux requires an acceptable 18\%. The fill factor remains reasonable and so this solution should supersede that of \citet{rgfz16}, although the implied 260\,km\,s$^{-1}$ space velocity is still unusually large for an MSP.

\begin{deluxetable*}{lrrrlrlr|rrrrr}[h!!]
\tablecaption{\label{Fits} Binary Model Fits}
\tablehead{
\colhead{Name} & \colhead{$i$}& \colhead{$f_c$}&  \colhead{$L_X/10^{33}$}&  \colhead{$T_N$}& \colhead{$d$}& $m_c$ &\colhead{$\chi^2/{\rm DoF}$} 
& \colhead{$f_i/f_*$} & \colhead{${\dot E}_c$} &\colhead{$\eta_{\rm heu}$}&\colhead{$\eta_\gamma$} & \colhead{$\eta_H$}\cr
& (deg)  & & (erg\,s$^{-1}$) & (K) & (kpc) &$^{(a)}$ &
& $^{(b)}$ & $^{(c)}$ & & &
}
\startdata
J0023+0923 & $77^{+13}_{-13}$ & $0.72^{+0.04}_{-0.04}$ & $2.96^{+0.16}_{-0.13}$ &$3340^{+50}_{-70}$ &$2.23^{+0.08}_{-0.08}$ &$1.8^{+0.2}_{-0.1}$& $335/62(5.41)$  
&0.46/0.48&1.15 & 0.29&0.18&0.12\cr  
J0251+2606 & 
$52.2^{+10.}_{-9.8}$ & $0.92^{+0.06}_{-0.09}$ & $2.52^{+0.12}_{-0.11}$ &$1140^{+600}_{-550}$ &$3.26^{+0.10}_{-0.10}$ &$3.2^{+0.6}_{-0.3}$& $72/32(2.25)$
&0.74/0.81&0.86& 0.34 & 0.47&0.23\cr     
J0636+5128 &$23.3^{+0.3}_{-0.4}$&$0.98^{+0.02}_{-0.02}$&$0.97^{+0.01}_{-0.01}$ &$1900^{+80}_{-90}$&$1.05^{+0.01}_{-0.01}$&$1.8^{+0.1}_{-0.1}$& $557/166(3.36) $
&$\infty$/0.56&0.56& 0.42&0.02&0.10\cr  
J0952$-$0607&$44.8^{+2.7}_{-2.2}$&$0.48^{+0.09}_{-0.10}$&$20.43^{+0.77}_{-0.76}$ &$2950^{+140}_{-170}$&$3.97^{+0.55}_{-0.67}$&$2.9^{+0.1}_{-0.1}$&$116/70(1.66)$ 
&1.16/0.89&6.65 & 0.12 &0.07&0.27\cr 

J1124$-$3653&$44.9^{+4.0}_{-1.9}$&$0.84^{+0.03}_{-0.03}$&$2.86^{+0.08}_{-0.10}$ &$1540^{+660}_{-710}$&$2.72^{+0.10}_{-0.08}$&$4.1^{+0.1}_{-0.3}$&56/40(1.40)
&1.16/0.99&0.52&0.44&2.48&0.54\cr  
J1301+0833  
&$44.0^{+2.8}_{-2.2}$&$0.68^{+0.06}_{-0.06}$&$6.97^{+0.46}_{-0.58}$ &$2430^{+90}_{-120}$&$2.23^{+0.08}_{-0.13}$&$4.5^{+0.1}_{-0.2}$& $161/126(1.28)$  
&1.00/0.89&2.53&0.20&0.15&0.18\cr 
J1959+2048  
&$62.5^{+1.3}_{-1.1}$&$0.90^{+0.01}_{-0.01}$&$30.10^{+0.36}_{-0.36}$ &$2670^{+30}_{-30}$&$2.04^{+0.01}_{-0.01}$&$3.6^{+0.1}_{-0.1}$& $140/89(1.57)$ 
&0.78/0.79&8.93 &0.11&0.07&0.27\cr 
J2052+1219&$54.5^{+2.9}_{-2.1}$&$0.99^{+0.01}_{-0.01}$&$9.44^{+0.24}_{-0.28}$ &$3020^{+90}_{-130}$&$3.94^{+0.07}_{-0.07}$&$4.2^{+0.1}_{-0.1}$& $121/84(1.44)$ 
&0.73/0.79&1.11&0.30&0.68&0.68\cr 
J2241$-$5236&$49.7^{+2.2}_{-1.9}$&$0.66^{+0.03}_{-0.04}$&$2.82^{+0.05}_{-0.06}$ &$2820^{+60}_{-60}$&$1.24^{+0.04}_{-0.05}$&$1.6^{+0.1}_{-0.1}$&163/62(2.64)
&0.87/0.86&1.77&0.24 &0.24&0.14\cr 
\enddata
\tablenotetext{(a)}{Companion mass in $10^{-2}M_\odot$, as inferred from the fit assuming $M_{\rm NS}=1.5\, M_\odot$ (except for J1301, J1959; see text).}
\tablenotetext{(b)}{``Outer gap'' (OG) conversion between the Earth line-of-sight ($f_i$, at inclination $i$) and equatorial plane ($f_*$) isotropic flux estimates and the corresponding true sky average. When $f_i/f_* >1$, the model predicts more companion heating than Earth-directed gamma-ray flux.}
\tablenotetext{(c)}{For pulsars with proper motion $\mu$ this spindown power (units of $10^{34}$\,erg\,s$^{-1}$) is Shklovskii-corrected for $I_{45}=1$ and our fit distance; see \S4.3. }

\end{deluxetable*}

{\bf PSR~J1959+2048} is the original BW MSP. Its pre-{\it Fermi} discovery likely stems from its relatively large $P_b=9.17$\,hr, since substantial time is thus spent out of eclipse with an unobscured pulsar detectable at radio wavelengths. Interest in inclination fitting is especially high because the \citet{vKBK11} measurement of a large companion radial velocity coupled with the \citet{reyet07} estimate of $i=65\pm2^\circ$ indicates a very high $M_p=2.40 \pm 0.12\, M_\odot$. We rely here on the optical/infrared photometry previously published by Reynolds et al.\ (2007). Refitting, we find a weak preference for high log\,$[Z]$ and a revised distance of 2.3\,kpc. The $i$ is consistent with, but even slightly smaller than, the R07 estimate. At this distance the direct-heating flux is 27\% of the corrected spindown power. Thus, like J1301, the revised fit is now consistent with direct heating, although the inferred heating power exceeds the $\gamma$-ray luminosity by a factor of $\sim 4$. This strengthens \citet{vKBK11}'s conclusion that, for direct heating, the neutron star mass is large, although this critical direct heating assumption may still be questioned, as we discuss in \citet{sr17} and in future work.

{\bf PSR~J2052+1219:} The MDM statistical errors seem to under-represent the true fluctuations, so these were doubled, dropping $\chi^2$ from 214 to 121. Otherwise the fit is quite good. For consistency with the other objects, we did not fit the two $u$ points, but they are close to the predicted model. The Keck $r$ and $i$ points are somewhat lower than the model, suggesting that the fit should have a higher $i$ and/or a lower $T_N$. Interestingly, this BW has the strongest preference for low log\,$[Z]$. In this case if $A_V$ is free it stays close to 0.3, independent of the model log\,$[Z]$. Spectra of this object may test the heavy-element abundance. The model fit implies a large distance quite consistent with the DM estimate. At this distance the inferred heating and $\gamma$-ray powers agree, but are a large fraction of the spin-down luminosity. 

{\bf PSR~J2241$-$5236} is particularly interesting since An et al. (2017) found evidence for GeV orbital modulation, suggesting that an intrabinary shock contributes in this band. The GeV peak is single, implying a tangent view of the shock, as would occur for $i\approx 30^\circ$--50$^\circ$, depending on the companion wind momentum. Our modest fit $i$ is consistent with this interpretation.

\subsection{Comparison of Efficiencies and Beaming Corrections}

Figure \ref{fOm} shows that for most magnetic inclinations $\alpha$ we expect little or no GeV flux at viewing angles $i< 30^\circ$. Thus J0636, whose shallow light curve strongly supports the fit low inclination, provides a basic test of the beaming picture. Indeed, it is an anomalously weak $\gamma$-ray pulsar (Table 1), only detected since the LAT events could be folded on the pre-existing radio MSP ephemeris. The heuristic luminosity law predicts a $\gamma$-ray efficiency of $\eta_{\rm heu}=0.42$; the observed 4FGL $\gamma$-ray flux limit corresponds to $\eta_{\rm obs} < 0.02$, so it is underluminous by a factor of $\sim 20$. Since we do not expect any OG flux directed at our fit $i = 23^\circ$, there is no meaningful $f_i$ in Table 3. (The flux we do see may be fainter GeV emission from lower altitude, directed along the magnetic pole.) Note that the required heating efficiency is only 10\% of $\eta_{\rm heu}$, which is easily accommodated.

However when we compare $\eta_\gamma$ and $\eta_H$ for our full sample, we find only a mild correlation (Spearman correlation $cor=0.35$). Clearly the beaming correction or the heating fit parameters are imperfect. Several factors can contribute toward this disagreement. The sizes of the efficiencies are an important clue. For example, for J1124 we have $\eta_\gamma = 2.5$. The moment of inertia may be larger than the standard $I_{45}=1$, which is plausible for heavy, highly recycled pulsars. But as noted above, an overestimate of the photometric distance likely dominates. Returning J1124 to its DM distance would decrease $\eta_\gamma$ to a large but plausible 0.4, which is in good agreement with $\eta_H$. This example highlights the fact that comparison of $\eta_\gamma$ and $\eta_H$, assuming an accurate beaming correction, gives an independent estimate of the distance. Such an estimate would decrease J0252's distance to $\sim 2.3$\,kpc, increase J0952's to $\sim 8$\,kpc, and increase J1959's to $\sim 4$\,kpc. However, it seems unlikely that distance errors are the full explanation. Beaming corrections, even if imperfect, are clearly needed for some BWs. Certainly J0636 requires such correction. Also, inspection of Figure \ref{fOm} shows that for our fit inclination $i=45^\circ$ for J0952, many $\alpha$ would not produce an OG GeV detection. Also, distance alone is unlikely to explain the disagreement for J1959, since at 4\,kpc the implied $\gamma$-ray luminosity would be $\sim 2.5$ times its expected heuristic level. Improved estimates of MSP beaming are therefore desirable.

In our present beaming estimates, we have been forced to average over $\alpha$ (horizontal green bands in Figure \ref{fOm}), which result in $a f_i>$ and $< f_\ast>$ close to unity, whereas corrections for particular angles can be quite substantial. Other observables might resolve this degeneracy. For example, the beaming model also predicts the phase separation $\Delta_\gamma$ of the principle $\gamma$-ray peaks. All of our BWs except J0023 and J0636 have a fairly well-defined double structure, so we could use this to constrain the $\alpha$ range (in our given inclination $i$ range) most consistent with the $\gamma$-ray pulse shape. However, the large $\Delta_\gamma$ of most of our BWs is only obtained in this OG model for $i>60^\circ$. Thus, if this OG model is applicable, the $i$ must be underestimated. Since this beaming model was computed for young pulsars rather than BWs, we might suspect that a pulse shape match will require a different model. Nevertheless, when applied to more sources and more modern MSP beaming maps, this direct view/radiative heating comparison provides a novel way to test $\gamma$-ray magnetosphere models. 

\subsection{Inclination Distribution}

\begin{figure}[h]
\includegraphics[width=9cm]{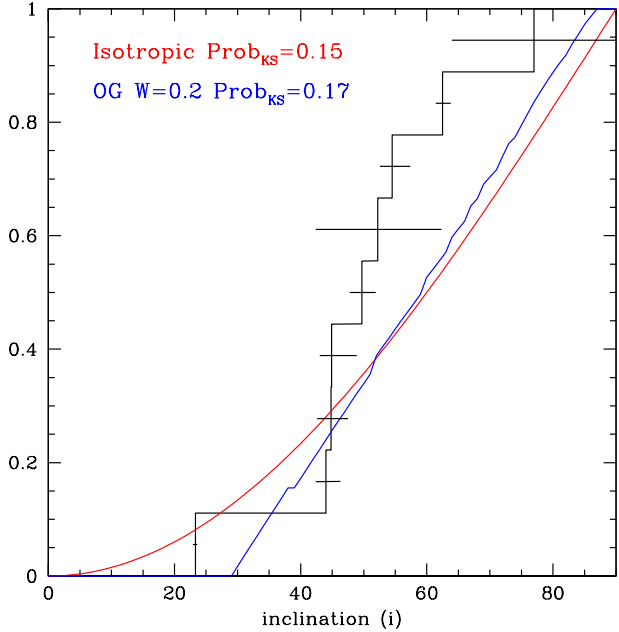}
\caption{\label{KS} Our fit inclination distribution compared with an isotropic distribution (red) and an OG model distribution with $\alpha$ isotropic and $w=0.2$ (blue).}
\end{figure}

We have seen that there is some tension between our fit $i$ values and the expectations of the OG model. Figure \ref{KS} compares the photometric fit $i$ values to a simple isotropic distribution and to the $i$ expected when one observes within the OG beam of Figure \ref{fOm}. In the OG case the lack of very low $i$ does boost somewhat the fraction expected to be viewed edge-on. The data seem to have an excess at $i=40$--60$^\circ$, but these differences are not statistically significant, with the KS test allowing both models. More objects will be needed to chose between these hypotheses.

If the mid-range $i$ excess above proves significant, it may be that BWs are most easily discovered at such inclinations. For example, the bright GeV MSP emission may be beamed away from the spin (orbital) axis, but companion evaporation produces a dense ionized plasma wind in the orbital plane, resulting in eclipses, dispersion, and scattering that will make it more difficult to detect BW radio counterparts at $i\approx 90^\circ$. The competition between these effects could bias detections to $i \approx 40^\circ$--60$^\circ$. 

\subsection{Companion Masses and Radii}

Taking our fit $i$ at face value, we can replace the commonly-used minimum companion masses with true masses. The resulting $m_c$, assuming a pulsar mass of 1.5\,$M_\odot$, are plotted with their uncertainty ranges in Figures 1 and 10. The masses cluster around $0.03\,M_\odot$ and it seems that all BW companions are heavier than $0.017\,M_\odot$. The biggest shift is, of course, for J0636, whose revised $M_c \approx 0.018\, M_\odot$ encroaches on the classical BW range.  

\begin{figure}[h]
\includegraphics[width=8.9cm]{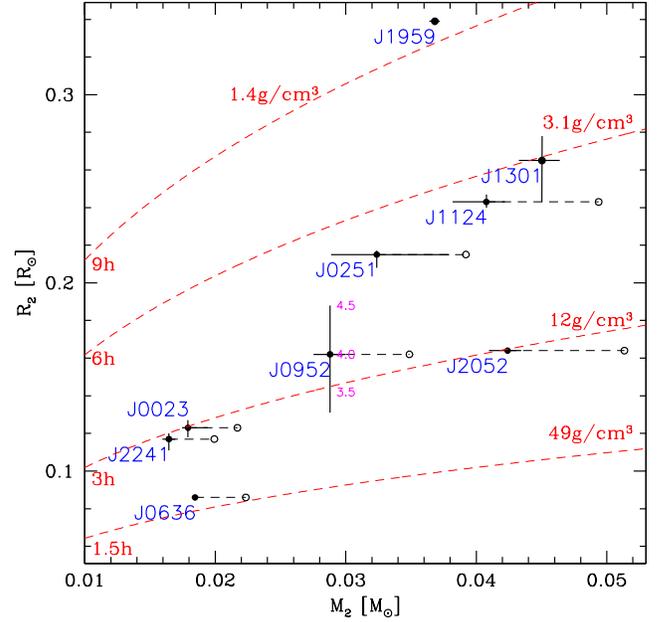}
\caption{\label{massrad} Companion masses and radii for an assumed $M_{\rm NS}=1.5\,M_\odot$ (solid dots) to $M_{\rm NS}=2\, M_\odot$ (open dots). For J1301 and J1959 we use the companion radial velocity to set the mass scale. For J0952 numbers on the $y$ error-bar show how the radius estimate correlates with distance in kpc.}
\end{figure}

With our inclination and fill factor estimates we can examine companion densities for clues to the evolution. First we recall the minimum density for a Roche-lobe-filling companion,
$$
\rho= 0.129 P_b^{-2}\, {\rm g\, cm^{-3}},
$$
with the orbital period in days. In Figure \ref{massrad} we show these lower density limits for several orbital periods. For comparison we plot our photometric estimates of the masses and radii. Here the error bars for the mass uncertainties incorporate only the $i$ measurement error and we assume $M_{\rm NS}=1.5\, M_\odot$. The dashed lines connect to the $M_{\rm NS}=2\,M_\odot$ position to display the mass-scale range. For J1301 and J1959 we can use the published companion radial velocity to set the mass scale. Since we wish to focus here on companion properties, we do not discuss the pulsar masses, but defer that to a future publication describing the spectroscopic modeling. 

Note that most companions are relatively close to Roche-lobe filling and hence near the minimum density for their orbital period. Displayed in this way, we see that the BWs in our sample seem to group into families; this appears as well in the $P_B$ distribution, and with a larger sample we can probe the evolution that leads to these objects. Note also that some objects are uncomfortably close to their minimum density for an assumed $M_{\rm NS}=1.5\,M_\odot$; this supports the conclusion of \citet{stret19} that RBs (and by extension BWs) are, as a class, massive NSs. J0952 is well within its Roche lobe at our best-fit photometric distance, and a direct-heating model has $\rho \approx 10\, {\rm g\,cm^{-3}}$, although with our photometric fill-factor/distance uncertainty the density may be as low as $6\,{\rm g\,cm^{-3}}$ or as high as $19\,{\rm g\,cm^{-3}}$. As noted above, the large best-fit distance for J1124 makes the inferred GeV efficiency unreasonably large at $\eta_\gamma=2.5$. This can be mitigated by smaller distances, at a cost in decreased fill factor and increased companion density. At its 1.1\,kpc DM distance we would find a companion density $\sim 97\,{\rm g\,cm^{-3}}$ (25$\times$ increase). This large density suggests evolution from Roche-lobe contact at $P_B\approx 0.9$\,hr, which seems implausible. Thus, while closer distances would help reduce the large $\eta_\gamma$, the distances are still likely to be larger than the DM estimate. A more substantial beaming correction $f_i$ could also help bring the efficiency estimates into agreement.

\section{Discussion and Conclusions}

We have collected a new set of photometric observations of BW MSPs and assembled multicolor light curves. We fit these data with a basic radiative heating model. On the whole the fits allow such direct pulsar irradiation, but there are some challenges to the model. First, in a number of cases the reduced $\chi^2$ is large. We suspect that underestimation of photometric errors is largely responsible, but a few objects (e.g., J0023) show evidence of nonthermal emission and others (e.g., J0251, J0952) may be variable. Such effects are not part of this basic model. The implied radiative heating efficiencies of 10\%--60\% are somewhat larger than assumed by other authors in the past. 

We also compare the required photon heating with the observed LAT GeV flux, which dominates the pulsar photon emission along Earth's line of sight. The estimated $\gamma$-ray efficiencies $\eta_\gamma$ are expected to be comparable to the ``heuristic'' $\gamma$-ray efficiency $\eta_{\rm heu}$, and to the (beaming-corrected) heating efficiency $\eta_H$ for a direct-heating model. The measured values do not always agree. For J0636 we can attribute the small $\eta_\gamma$ to strong beaming away from Earth. But for J1124, the very large inferred $\eta_\gamma$ probably indicates an overestimate of $d$. Conversely, for J0952 and J1959 the fit companion heating is more than twice the observed GeV luminosity at the photometric fit distance.  While the inferred GeV flux would increase with $d$, this would rapidly exceed the expected heuristic gamma-ray luminosity. So for these objects a larger beaming correction (gamma-rays more tightly beamed at the companion) seems to be a more likely solution. Interestingly, the observed $\gamma$-ray pulse widths suggest larger $i$ for individual objects in this model. This may indicate some bias in the direct-heating $i$ fits, but it would be useful to also compare with the predictions of other gamma-ray beaming models.

The fits also give interesting information on the companions, which substantially fill their Roche lobes, with companion densities moderately larger than the minimum from their orbital period. Some objects were fit with substantially larger distance than the DM estimates (J0251, J1124). As noted above, these require unrealistically large $\gamma$-ray fluxes (unless the $\gamma$-rays are tightly beamed toward Earth, as can happen at larger $i$). Conversely, if the DM distances are adopted, the photometry demands small companion radii and large companion densities for these two objects, an evolutionary challenge.

The light-curve shapes also hint that other physical processes contribute to the optical emission. Small phase shifts in the optical maximum can be seen (e.g., J0636), and a few have maxima broader than the best-fit model curves in some bands (e.g., J1124) or have blue color minima deeper than the models (e.g., J2052). All of these suggest that some additional heat source or heat transport mechanism may be present. Since the distribution of measured $i$ values is somewhat weighted to intermediate latitudes, and since the $\gamma$-ray pulse shapes generally prefer $i$ larger than found in our fits, we may suspect that these effects introduce some bias into the measurements. It is important to continue to search for such effects since decreased $i$ result in increased $M_{\rm NS}$ measurements for these BWs. As these are among the heaviest neutron stars known, this is important for equation-of-state constraints. 

Overall, our photometry shows that the observed GeV flux is a plausible source of the direct heating of BW companions. But this is true only if the photometric distance estimates are substantially wrong in several cases. Since the fit distance also affects the companion size and density estimates and, less directly, the inferred system inclination, it is important to eliminate distance errors. We can of course continue to address the source of companion heating with improved multicolor photometry and spectroscopy, but it seems as if robust independent distances, ideally from pulsar timing or VLBI parallax measurements, will be an important tool in understanding BW heating.
\bigskip
\bigskip

We thank D. Kandel, who helped with the ICARUS code, and the referee, whose careful reading and detailed comments helped us improve the paper. We are also grateful to Julia Deneva and Thankful Cromartie, who shared updated ephemerides for J0251 and J2052 in advance of publication. Daniel Perley and Brad Cenko kindly allowed us to use (and assisted with) their Keck LRIS imaging and spectroscopy reduction pipelines.

We are grateful to the staff at the Keck Observatory for their assistance. The W. M. Keck Observatory is operated as a scientific partnership among the California Institute of Technology, the University of California, and the National Aeronautics and Space Administration (NASA); it was made possible by the generous financial support of the W. M. Keck Foundation. We extend special gratitude to those of Hawaiian ancestry on whose sacred mountain we are privileged to be guests. 
This work is based in part on observations obtained at the MDM Observatory, operated by Dartmouth College, Columbia University, Ohio State University, Ohio University, and the University of Michigan.

R.W.R. was supported in part by NASA grant 80NSSC17K0024. 
The research of A.V.F.ʼs group was supported by the Christopher R. Redlich Fund, the TABASGO Foundation, and the Miller Institute for Basic Research in Science (U.C. Berkeley). 

\bigskip
\noindent{{\it Software:} ICARUS \citep{bet12,rs16}, IRAF}

\end{document}